\documentclass[12pt]{article}

\usepackage{amsmath}
\usepackage{amssymb}
\usepackage{amsthm}

\usepackage{upref}

\newtheorem{thm}{Theorem}[section]

\newcommand\Cal[1]{{\cal #1}}

\newcommand\reals{\mathbb R}
\newcommand\naturals{\mathbb N}
\newcommand\complex{\mathbb C}
\newcommand\integers{\mathbb Z}

\newcommand\cld{{\rm wor}}
\newcommand\clr{{\rm ran}}
\newcommand\nran{n^{\rm ran}(\e)}
\newcommand\nwor{n^{\rm wor}(\e)}

\newcommand\ket[1]{|#1 \rangle}

\newcommand\w{{\rm quant}}

\newcommand\comp{{\rm comp}}
\newcommand\compwor{{\rm comp}^{{\rm wor}}}
\newcommand\compran{{\rm comp}^{{\rm ran}}}
\newcommand\costwor{{\rm cost}^{{\rm wor}}}
\newcommand\costran{{\rm cost}^{{\rm ran}}}
\newcommand\cc{{\bf c}}

\def\c{{\gamma}}

\def\a{{\alpha}}
\def\e{{\varepsilon}}
\def\l{{\lambda}}
\def\E{{\mathbb E}}
\def\L{{\mathbb L}}
\def\Q{{\bf Q}}
  
\def\cb{\mathrm{comp}^{\mathrm{qub}}}

\title{\bf Classical and Quantum Complexity\\
of the Sturm-Liouville Eigenvalue Problem}
\author{A. Papageorgiou$^1$ and H. Wo\'zniakowski$^2$ \\
{\small $^{1,2}$Department of Computer Science, Columbia University, 
New York, USA} \\
{\small $^2$Institute of Applied Mathematics and Mechanics, 
University of Warsaw, Poland}}

\date{\today}
\begin{document}

\setcounter{page}{1}
\maketitle

\begin{abstract} 
We study the approximation of the smallest eigenvalue of a Sturm-Liouville 
problem in the classical and quantum settings. We consider a 
univariate Sturm-Liouville eigenvalue problem with a nonnegative function 
$q$ from the class $C^2([0,1])$ and study the minimal number $n(\e)$ of 
function evaluations or queries that are necessary to compute 
an $\e$-approximation of the smallest eigenvalue. We prove that 
$n(\e)=\Theta(\e^{-1/2})$ in the (deterministic) worst case setting, and 
$n(\e)=\Theta(\e^{-2/5})$ in the randomized setting. 
The quantum setting offers a polynomial speedup with {\it bit} queries and 
an exponential speedup with {\it power} queries.
Bit queries are similar to the oracle calls used in Grover's algorithm 
appropriately extended to real valued functions. Power queries are used for 
a number of problems including phase estimation. They are obtained by 
considering the propagator of the discretized system at a number of different 
time moments. They allow us to use powers of the unitary matrix 
$\exp(\tfrac12 {\rm i}M)$, where $M$ is an $n\times n$ matrix obtained from 
the standard discretization of the Sturm-Liouville differential operator. 
The quantum implementation 
of power queries by a number of elementary quantum gates that is polylog 
in $n$ is an open issue.

In particular, we show how to compute
an $\e$-approximation with probability $\frac34$ using 
$n(\e)=\Theta(\e^{-1/3})$ bit queries. For power queries, 
we use the phase estimation algorithm as a basic tool
and present the algorithm that solves the problem 
using $n(\e)=\Theta(\log\e^{-1})$ power queries, 
$\log^2\e^{-1}$ quantum operations,
and $\frac32 \log \e^{-1}$ quantum bits. We also prove that the minimal number
of qubits needed for this problem (regardless of the kind of queries used) 
is at least roughly $\tfrac12 \log \e^{-1}$. The lower bound on the number 
of quantum queries is proven in \cite{Bessen}. 

We derive a formula that relates the Sturm-Liouville eigenvalue problem 
to a weighted integration problem. Many computational problems may be recast 
as this weighted integration problem, which allows us to solve them with 
a polylog number of power queries. Examples include Grover's search, the 
approximation of the Boolean mean, NP-complete problems, and many multivariate
integration problems. In this paper we only provide the relationship formula. 
The implications are covered in \cite{PW04}.

\end{abstract}

\vskip 2pc
\section{Introduction}
\vskip 1pc
The study of the potential power of quantum computers has been a major
theoretical challenge. There will be an additional
incentive to build a quantum computer if we can identify computationally
important problems for which quantum computation offers
significant speedups over computation on a classical computer. 

For discrete  problems, the best known quantum algorithms
are due to Shor and Grover, see \cite{shor,grover}.
Shor's algorithm for factorization has an
exponential speedup over all {\it known} algorithms on a classical computer. 
Still, we can not yet claim that we have an exponential speedup
for this problem, since the complexity of factorization on a classical
computer is unknown. Grover's algorithm for data search offers a
quadratic speedup. 

For continuous problems, quantum complexity is known for linear
problems such as  
multivariate integration, path integration and multivariate
approximation,  see \cite{heinrich,H03,H04a,H04b,N01,TW02}. For these problems
we have an exponential speedup over the worst case setting, and
a  polynomial speedup over the randomized setting. The first quantum
study of a nonlinear continuous problem was done in \cite{Kacewicz} for
ordinary differential equations with polynomial speedups over the
classical settings.  

The purpose of this paper is to present classical and quantum 
complexity results of another nonlinear continuous problem. This
continuous problem is quite natural and computationally important, since it
corresponds to the (simplified) univariate Sturm-Liouville 
eigenvalue problem. The Sturm-Liouville eigenvalue problem is defined in 
\cite{courant} in full generality.
Here it is defined as finding 
the smallest eigenvalue of the differential operator
$$
\L_qu\,(x)\,:=\,-u^{\prime\prime}(x)\,+\,q(x)\,u(x)\qquad\mbox{for}\ \ 
 x\in (0,1),
$$
with the boundary conditions $u(0)=u(1)=0$. We assume that
the function $q$ is non-negative and belongs to the class $C^2([0,1])$ of twice
continuously differentiable functions whose norm
$\|q\|:=\max_{i=0,1,2}\max_{x\in [0,1]}|q^{(i)}(x)|$ is bounded by $1$.  
The operator $\L_q$ maps $C^2([0,1])$ into $C([0,1])$. 

The Sturm-Liouville eigenvalue problem has been extensively studied 
in the literature. The properties of the eigenvalues and the
eigenfunctions are well known and so are numerical algorithms 
for approximating them on a classical computer, see, 
e.g. \cite{babuska,collatz,courant,keller,strang}. 
Nevertheless, the complexity of approximating the smallest eigenvalue 
in the worst case and randomized settings, as well as
in the quantum setting, has not yet been addressed.

In this paper we study classical and quantum algorithms.
We prove bounds on the worst case and randomized 
complexities on a classical computer, and bounds on the 
query complexity and on the qubit complexity. 
We prove that
the complexity in the classical settings is a polynomial in
$\e^{-1}$. 

We study the quantum setting with {\em bit} queries and prove
polynomial speedups over the classical settings. Bit queries
correspond to approximate computation of function values, see~\cite{heinrich}, 
and are used in all papers dealing with the quantum study
of continuous problems.

We also study the quantum setting with {\em power} queries. 
Such queries are formally defined in Section 5.2. Here we only mention
that they are used in the phase estimation algorithm, which is the core
of many quantum algorithms including Shor's and Grover's algorithms.
Power queries are controlled-$W^{p_j}$ queries for some $n\times n$
unitary matrix $W$ and some exponents~$p_j$. For the phase estimation
algorithm, we have $p_j=2^{j-1}$ for  $j=1,2,\dots,m$, with~$m$ of
order $\log\,\e^{-1}$. 
For the factoring problem of a large integer $N$, Shor's algorithm
uses the unitary matrix $W$ such that power queries can be implemented 
by at most $O(\log^3N)$ elementary quantum gates.

For the Sturm-Liouville eigenvalue problem, as well for all problems 
studied in \cite{PW04}, we use power queries with the specific unitary
matrix
\begin{equation}\label{matrixW}
W\,=\,\exp\left({\tfrac12\,\mathrm{i}\,M_q}\right)\qquad \mbox{with}\qquad
\mathrm{i}\,=\,\sqrt{-1},
\end{equation}
where $M_q$ is an $n\times n$ real symmetric
tridiagonal matrix that is a classical approximation of the differential 
operator $\L_q$, see Section 3.2. The matrix $M_q$ depends on the
values of $q(j/(n+1))$ that appear on the diagonal of $M_q$ for
$j=1,2,\dots,n$. 

Unitary matrices similar to (\ref{matrixW}) play a key role
in quantum mechanics. They give the solution of the Schr\"odinger
equation, they are the propagator of a system evolving with 
Hamiltonian $M_q$, and are important in quantum simulation, 
see  \cite{nielsen}. Zalka \cite{zalka} deals with their
implementation. 
The crucial point about power queries is that we can use  
$W^j$ of the matrix $W$ given by (\ref{matrixW}) as one quantum query
for some $j$. Hence, lower bound results for bit queries do not apply to 
power queries.

We prove that in the quantum setting with power queries,
the Sturm-Liouville eigenvalue problem requires only roughly $\log\,\e^{-1}$ 
power queries with the matrix $W$ of (\ref{matrixW}). 
As shown in \cite{PW04}, many computational problems
can be reduced to the solution of the Sturm-Liouville eigenvalue
problem, and they can be also solved in 
polylog number of power queries. The list of such
problems include Grover's search, NP-complete problems,
and many continuous problems. This proves that the quantum setting
with power queries with the matrix $W$ of (\ref{matrixW}) 
is exponentially more powerful than the quantum setting with
bit queries. 

We stress that, contrary to Shor's algorithm, we do {\em not} know
if power queries with the $n\times n$ matrix $W$ of (\ref{matrixW}) can be
implemented by a number of existing elementary quantum gates that is polylog 
in $n$. We asked a number of colleagues and most of them doubt whether
this can be achieved. If this is indeed the case, then the positive
results on the polylog number of such power queries will be of only  
theoretical interest. Still, if a future quantum computer is
able to perform such power queries in a polylog number of, 
perhaps, more general elementary quantum gates or by some other
quantum devices, the
polylog number of power queries will lead to efficient quantum
algorithms, and will allow us to solve many computational problems 
exponentially faster than on a classical computer. {}From this point of view, 
we may interpret the positive results on the number of power queries
with the matrix~$W$ of (\ref{matrixW}) as the indication that 
building a quantum computer with such queries would be a very
desirable task which would give us a very powerful computational device.

\section{Survey of the Results}

In this section we explain our results in more technical terms. 
For a classical computer, we study the worst case and randomized
settings in the real number model of computation with oracles,
see \cite{novak,traub,TW98}. That is, we assume that arithmetic operations
(addition, subtraction, multiplication, division, and evaluation of 
elementary functions), as well as comparisons of real numbers, are 
performed exactly with cost taken as unity. We also assume that 
the information about functions $q$ is given by sampling $q$ 
at finitely many points with the cost of one function
evaluation taken as $\cc$. Typically $\cc\gg 1$.

We want to approximate the smallest eigenvalue $\l(q)$ of the operator
$\L_q$  to within $\e$. Let $n(\e)$ be the smallest number of function 
values of $q$ needed to compute such an $\e$-approximation in a given
setting. The number $n(\e)$ is called the {\em information complexity}.
The {\em complexity}, $\comp(\e)$, is defined as 
the minimal total cost of computing an
$\e$-approximation in a given setting. Obviously we have
$$
\cc\,n(\e)\,\le\,\comp(\e).
$$

We prove that in both classical settings, the complexity of the
Sturm-Liouville eigenvalue problem is polynomial in $\e^{-1}$, or
equivalently is exponential in the number $\lfloor \log\,\e^{-1}\rfloor $ 
of correct bits of a computed approximation.
More precisely, there exist positive numbers $\a_i$ independent of
$\e$ such that:
\begin{itemize}
\item in the worst case setting,
\begin{eqnarray*}
\a_1\,\e^{-1/2}\,\le\,&n(\e)&\,\le\,\a_2\,\e^{-1/2},\\
\a_1\,\cc\,\e^{-1/2}\,\le\,&\comp(\e)&\,\le\,
\a_2\,\cc\,\e^{-1/2}\,+\,\a_3\,\e^{-1/2}\,\log\,\e^{-1},
\end{eqnarray*}

\item in the randomized setting,
\begin{eqnarray*}
\a_4\,\e^{-2/5}\,\le\,&n(\e)&\,\le\,\a_5\,\e^{-2/5},\\
\a_4\,\cc\,\e^{-2/5}\,\le\,&\comp(\e)&\,\le\,\a_5\,\cc\,\e^{-2/5}\,+\,
\a_6\,\e^{-1/2}\,\log\,\e^{-1}.
\end{eqnarray*}
\end{itemize}

The lower bounds on $n(\e)$, and consequently on $\comp(\e)$, 
are obtained by relating the 
eigenvalue problem to the integration problem for functions from the
unit ball of $C^2([0,1])$. It is well known that the minimal number 
of function values for this integration problem is bounded from below
by roughly $\e^{-1/2}\,$ in the worst case setting and by $\e^{-2/5}$ in the
randomized setting; see, e.g., \cite{novak,traub} and the survey of
these results in \cite{TW98}.
 
The upper bounds on $n(\e)$ and $\comp(\e)$ in the worst case setting
are obtained by the cost of the classical algorithm that computes an
$\e$-approximation by the bisection algorithm of the Sturm
sequence \cite[p.~300]{wilkinson}, see also \cite[Ch.~5.3.4]{demmel},
applied to an $n\times n$ matrix which is the classical
discretization of the operator $\L_q$ with $n=\Theta(\e^{-1/2})$.
The matrix depends on $n$ function values of $q$ computed at
equidistant points of $[0,1]$. Since we need roughly $\log\,\e^{-1}$ 
bisection steps, and the cost of each step is proportional to $n$, 
the total cost 
is of order $(\cc+\log\,\e^{-1})\e^{-1/2}$. Hence, modulo the logarithm
of $\e^{-1}$, the worst case complexity is of order $\cc\,\e^{-1/2}$.  

The upper bounds on $n(\e)$ and $\comp(\e)$ in the randomized setting
are obtained by the following algorithm. We first approximate the
function $q$ by a natural cubic spline $\bar q$ using 
$n$ deterministic sample points of $q$ at equidistant points 
of $[0,1]$ with $n=\Theta(\e^{-2/5})$. The relationship
between the smallest eigenvalue and integration problems, see Section~3,
states that  
\begin{equation}\label{1}
\l(q)\,=\,\l(\bar q)\,+\,\int_0^1\left(q(x)-\bar q(x)\right)
u_{\bar q}^2(x)\,dx
\,+\,O(n^{-4}).
\end{equation}
Here $u_{\bar q}$ is the normalized eigenfunction, $\int_0^1u_{\bar
q}^2(x)\,dx=1$, corresponding to the smallest eigenvalue $\l(\bar q)$. 

Since we have complete information on the spline $\bar q$, we may 
approximate $\l(\bar q)$ and $u_{\bar q}$ with arbitrarily small error. 
For $\l(\bar q)$, we achieve an error of order $\e$ 
as in the worst case setting, with cost proportional to 
$\e^{-1/2}\log\,\e^{-1}$. To obtain an approximation to $u_{\bar q}$, 
we apply one step of the inverse power algorithm with an appropriately
chosen initial vector. In this way we obtain a vector, from which we compute 
$u_{\bar q}$ via piecewise interpolation. 
The total cost of computing $\l(\bar q)$ and $u_{\bar q}$ is of order 
$\e^{-1/2}\log\,\e^{-1}$. 

We then approximate the second term in (\ref{1}) using 
the Monte Carlo algorithm for the function 
$(q(x)-\bar q(x))u_{\bar q}^2(x)$ computed at $n$ randomized points
with uniform distribution over $[0,1]$. This leads to an
$\e$-approximation in the randomized setting with cost 
bounded from above by a quantity proportional to
$\cc\,\e^{-2/5}+\e^{-1/2}\log\,\e^{-1}$, where the first term 
bounds the information cost and the second term bounds the combinatorial 
cost of the algorithm. Hence, we have a sharp
estimate on the randomized information complexity $n(\e)$.  
The ratio of the upper to lower bounds
of the randomized complexity is roughly at most $\e^{-1/10}$. 

In both classical settings, algorithms for which we obtain
upper bounds on complexity require space of order $\e^{-1/2}$.
This follows from the fact that we need to work on $n\times n$ tridiagonal
matrices with $n$ of order $\e^{-1/2}$.

We now turn to the quantum setting. Quantum algorithms are described
in Section 4. Here we only mention that quantum algorithms
work on $2^\nu\times 2^\nu$ unitary matrices, where $\nu$ is  
the number of qubits. The qubit complexity is defined as the minimal 
number of qubits needed to solve a problem. Roughly speaking, the
qubit complexity corresponds to the space complexity for a classical computer.
For the foreseeable future, qubits will be a scarce 
resource. That is why the qubit complexity is especially important, 
and computationally important problems with relatively small qubit 
complexity are of special interest. 

We prove that the qubit complexity, $\cb(\e)$, of the 
Sturm-Liouville eigenvalue problem is
of order $\log\,\e^{-1}$, which is relatively modest.
In this paper $\log$ denotes $\log_2$. More precisely, we prove that
$$
\tfrac12\,\log\,\e^{-1}\,+\,\Omega(1)\,\le\,\cb(\e)\,\le\,
\tfrac32\,\log\,\e^{-1}\,+\,O(1).
$$
These bounds hold regardless of the kind of queries used.
Clearly, the qubit complexity yields a lower bound for the cost of any quantum
algorithm solving this problem.

We now turn to the quantum setting with bit queries.
We show that the bit query complexity is
$\Theta(\e^{-1/3})$.
This result is obtained by using:
\begin{itemize}
\item  equation (\ref{1}) relating the
Sturm-Liouville eigenvalue problem to integration, 
\item a lower bound on bit queries for integration, and 
\item a  modification of the classical randomized 
algorithm described above that uses a quantum summation algorithm 
instead of Monte Carlo to approximate the weighted integral 
in (\ref{1}).
\end{itemize}

We now discuss the quantum setting with power queries. In this setting,
the Sturm-Liouville eigenvalue problem can be solved using 
the well-known phase estimation algorithm as a basic
tool, see, e.g., \cite[Section~5.2]{nielsen}. This algorithm uses
power queries and the quantum inverse Fourier transform as its main
ingredients. The power queries have the form controlled-$W^{2^j}$ for
$j\in\naturals$, i.e., they use powers of the matrix
$W=\exp\left(\frac12\mathrm{i}\,M_q\right)$, with $M_q$ an $n\times
n$ real symmetric tridiagonal matrix whose diagonal 
elements depend on the values of~$q$.
The matrix $M_q$ is a well-known discretization of the differential
operator $\L_q$, and its size $n$ depends on the necessary accuracy.
To obtain an $\e$-approximation we use $n$ of order $\e^{-1/2}$. 

The phase estimation algorithm uses the exact
eigenvector of $M_q$, equivalently of $W$,  
as part of its initial state, see \cite[Section~5.2]{nielsen}. Abrams and
Lloyd \cite{abrams} analyzed the case when the exact eigenvector
is replaced by an approximate eigenvector and concluded that as long
as the approximation is {\em good enough}, 
the phase estimation algorithm will still supply a good
approximation to the corresponding eigenvalue. Jaksch and Papageorgiou
\cite{jaksch} proposed an efficient construction of an approximate eigenvector.
Their idea was to solve the problem with low accuracy on a classical 
computer and obtain a \lq\lq short\rq\rq vector which approximates 
the eigenfunction $u_q$ at few points. Then the amplitudes of this
short vector are replicated on a quantum computer by the Hadamard
transform, which yields a \lq\lq long\rq\rq (vector) state that can be used
as the approximate initial state in the phase estimation algorithm.

We show how the construction of Jaksch and Papageorgiou
can be used for the Sturm-Liouville eigenvalue problem. In this way,
we compute an $\e$-approximation of the smallest eigenvalue with
probability~$\tfrac34$ by the phase estimation algorithm using
$\log\,\e^{-1}\,+O(1)$ power queries. 
The algorithm requires an additional number of quantum operations
at most of order $\log^2\e^{-1}$. This additional cost 
is for the quantum inverse Fourier transform. 
Finally, the number of qubits is
$\frac32\,\log\,\e^{-1}\,+\,O(1)$.
A lower bound on the number of power queries of order $\log\,\e^{-1}$ 
has been proven in \cite{Bessen}. 

Comparing these quantum estimates to the classical complexity bounds in the
worst case and randomized setting, we see that the quantum setting
with power queries yields an exponential speedup between the
number of power queries and the number of function values needed 
for the Sturm-Liouville eigenvalue problem. 

Finally, we point out  important consequences of our results,
which we study in detail in \cite{PW04}. 
Knowing that the Sturm-Liouville eigenvalue problem
can be solved with polylog power queries, it is natural
to study which computational problems can be reduced to this problem.
In this respect, we think that the most important result of this paper 
is the formula that relates this eigenvalue problem to integration. 
In a particular case, this formula, see~(\ref{333}), states that 
\begin{equation}\label{444}
\l(q)\,=\,\pi^2+\tfrac12\,+\,2\int_0^1\left(q(x)-\tfrac12\right)
\sin^2(\pi x)\, dx\,+\,O\left(\|q-\tfrac12\|_{\infty}^2\right).
\end{equation}
Hence, the problem of computing the smallest eigenvalue is equivalent,
modulo the second order term, to the weighted integration problem.
Since $\l(q)$ can be approximated with polylog power queries, so can 
the weighted integral of $q$. It turns out that many computational 
problems can be formulated as an integration problem. Examples include
important discrete problems such as Grover's search, the
approximation of the Boolean mean, and NP-complete problems.
The approximation of the Boolean mean is used as the
primary tool to compute multivariate integrals and path
integrals. Hence, all these problems can be solved by 
reducing them to the Sturm-Liouville 
eigenvalue problem with a polylog number of 
power queries in the quantum setting. 
It is well-known that Grover's search and the approximation of the Boolean mean
require a number of bit queries polynomial in the problem size, 
which in our case is a polynomial in $\e^{-1}$.
This shows that power queries are exponentially more powerful than
bit queries, see \cite{PW04} for details. 

\vskip 2pc
\section{Problem Definition}
We deal with functions from the class
$$
\Q\,=\,\big\{\,q:[0,1]\to [0,1]\ \big|\ \ q\in C^2([0,1])\ \
\mbox{and}\ \ 
\|q\|:=\max_{i=0,1,2}\ \max_{x\in [0,1]}|q^{(i)}(x)|\,\le 1\,\big\}. 
$$
For a function $q\in \Q$, we consider the Sturm-Liouville eigenvalue problem 
$\L_qu=\l\,u$ for a non-zero $u$, or equivalently
\begin{equation}
u^{\prime\prime}(x) - q(x) u(x) + \lambda u(x) = 0, 
\quad {\rm for\ } \ x\,\in\,(0,1),
\label{eq:SL1}
\end{equation}
with the boundary conditions
\begin{equation}\label{eq:SL2}
u(0)=u(1)=0.
\end{equation}
Let $\l=\l(q)$ be the smallest eigenvalue of 
(\ref{eq:SL1}), (\ref{eq:SL2}). Multiplying (\ref{eq:SL1}) by $u$
and integrating by parts, see \cite{babuska,courant,strang}, 
we conclude that the smallest eigenvalue satisfies 
\begin{equation}\label{var}
\l(q)\,=\,\min_{0\ne u\in H_0^1} \frac{\int_0^1\left[ (u^\prime(x))^2 + q(x)
u^2(x)\right] \, dx}
{\int_0^1 u^2(x)\, dx},
\end{equation}
where $H_0^1$ is the Sobolev space of absolutely continuous\footnote{
A function $f$ is absolutely continuous if and only if it can be written as 
$f(x)=f(0)+\int_0^xf'(t)dt$ for all $x\in[0,1]$.} functions
for which $u^\prime\, \in L_2([0,1])$ and $u(0)=u(1)=0$.

Let $u_q$ be a normalized real eigenfunction corresponding to the smallest
eigenvalue. It is known that the eigenvalues of $\L_q$ are simple, and
the eigenspace corresponding to $\l(q)$ is of dimension one. Therefore 
$u_q$ is uniquely defined up to the sign. In particular, $u_q^2$ is
uniquely defined. Then (\ref{var}) states that
\begin{equation}
\l(q)\,=\,\int_0^1\left(\left(u^{\prime}_q(x)\right)^2\,+\,q(x)u^2_q(x)\right)
\, dx\qquad\mbox{and}\qquad \|u_q\|_{L_2}\,:=\,\left(
\int_0^1u_q^2(x)\,dx\right)^{1/2}\,=\,1.
\end{equation}

Observe that $q\in \Q$ implies that $u_q\in C^4([0,1])$.
Since $\|q\|\le 1$, and $\|u_q\|_{L_2}=1$ with $u_q(0)=u_q(1)=0$,
then $|u^{(i)}_q(x)|$ are uniformly bounded for all $i\in[0,4]$,
$x\in [0,1]$ and $q\in \Q$, see e.g., \cite[p. 337]{courant}. 

The smallest eigenvalue $\l(q)$ is a non-decreasing function of $q$,
i.e., $q_1(x)\le q_2(x)$ for $x\in [0,1]$ implies $\l(q_1)\le
\l(q_2)$. It is known that for $q\equiv c$ we have 
$$
\l(c)\,=\,\pi^2+c\qquad\mbox{and}\qquad u_c(x)\,=\,\sqrt{2}\,\sin(\pi
x).
$$
This implies that for $q\in \Q$, we have $\l(q)\in
[\l(0),\l(1)]=[\pi^2,\pi^2+1]$. 

We will need estimates of the smallest eigenvalues and their
eigenfunctions for perturbed functions $q$. This is a classical
problem and many such estimates can be found in the literature,
not only for the simplified Sturm-Liouville problem that we consider in
this paper but also for more general eigenvalue problems. In our
case, the problem of perturbed eigenvalues and eigenvectors
is well-conditioned, since the differential operator
$\L_q$ is symmetric and  the eigenvalues of $\L_q$ are well separated.   
Combining results from \cite{courant,keller,titchmarsh} one can
obtain the following estimates for $q,\bar q\in \Q$:
\begin{eqnarray}
|\l(q)-\l(\bar q)|\,&\le&\,\|q-\bar q\|_{\infty}\,:=\,\max_{x\in
 [0,1]}|q(x)-\bar q(x)|,\label{111}\\
\|u_q-u_{\bar q}\|_{\infty}\,&=&\,O\left(\|q-\bar q\|_{\infty}
 \right),\label{222}\\
\l(q)\,&=&\,\l(\bar q)\,+\,\int_0^1\left(q(x)-\bar q(x)\right)
u_{\bar q}^2(x)\,dx
\,+\, O\left(\|q-\bar q\|_{\infty}^2\right).\label{333}
\end{eqnarray} 
We stress that the factors in the big-$O$ notation are independent
of $q$ and $\bar q$.

These relations follow by elementary arguments. Indeed, (\ref{111})
follows from (\ref{var}) by taking $u=u_{\bar q}$, which leads 
to $\l(q)-\l(\bar q)\le \|q-\bar q\|_{\infty}$. By replacing the roles
of $q$ and $\bar q$ we get $\l(\bar q)-\l(q)\le \|q-\bar
q\|_{\infty}$, which implies (\ref{111}). The next relation (\ref{222})
can be also proved by a matrix approximation to the operator $\L_q$,
which will be done in Section~4. Finally, (\ref{333}) follows by 
again taking $u=u_{\bar q}$ in (\ref{var}), which leads to 
\begin{eqnarray*}
\l(q)\,&\le&\,\l(\bar q)+
\int_0^1\left(q(x)-\bar q(x)\right)\,u_{\bar q}^2(x)\,dx \\
&=&\,\l(\bar q)+
\int_0^1\left(q(x)-\bar q(x)\right)\,u_{q}^2(x)\,dx +
\int_0^1\left(q(x)-\bar q(x)\right)\,\left(u_{\bar
q}^2(x)-u_q^2(x)\right)\,dx.
\end{eqnarray*}
By (\ref{222}), the last term is of order $\|q-\bar q\|^2_{\infty}$.
Taking $u=u_q$ in the expression (\ref{var}) defining $\l(\bar q)$, we obtain
$$
\l(\bar q)\,\le\,\l(q)+\int_0^1\left(\bar q(x)-q(x)\right)u_q^2(x)\,dx.
$$
The last two inequalities imply (\ref{333}). We shall see later that
the formula (\ref{333}) will be very useful in deriving lower bounds
for classical algorithms. Note that if we take $\bar q\equiv \tfrac12$, then
the formula (\ref{333}) becomes (\ref{444}).

\section{Classical Algorithms}
In this section we consider classical algorithms, i.e.,
algorithms on a classical (non-quantum) computer. These algorithms
can be either deterministic or randomized. They  use information
about the functions $q$ from $\Q$ by computing $q(t_i)$ for
some discretization points $t_i\in [0,1]$.
Here, $i=1,2,\dots,n_q$, for some $n_q$, and the points $t_i$  
can be adaptively chosen, i.e., $t_i$ can be a function
$$
t_i\,=\,t_i(t_1,q(t_1),\dots,t_{i-1},q(t_{i-1})),
$$
of the previously computed function values and points
for $i\ge 2$. The number $n_q$ can also be adaptively chosen, 
see, e.g., \cite{traub} for details. 

A classical deterministic algorithm produces an approximation
$$
\phi(q)\,=\,\phi (q(t_1),\dots,q(t_{n_q}))
$$ 
to the smallest eigenvalue $\l(q)$ based on 
finitely many values of $q$ computed at deterministic points.
Let $n=\sup_{q\in \Q}n_q$. We assume that $n<\infty$. 
The worst case error of such a deterministic algorithm $\phi$ 
is given by
\begin{equation}
e^{\cld}(\phi,n) = 
\sup_{q\in \Q}|\l(q) - \phi(q)|. 
\label{eq:cde}
\end{equation}

A classical randomized algorithm produces an approximation to $\l(q)$
based on finitely many values of $q$ computed at random points, and
is of the form 
$$
\phi_{\omega}(q)\,=\,\phi_{\omega}(q(t_{1,\omega}),
\dots,q(t_{n_{q,\omega},\omega})), 
$$
where $\phi_\omega,t_{i,\omega}$ and $n_{q,\omega}$ are random 
variables. We assume that the mappings
\begin{eqnarray}
\omega &\mapsto& t_{i,\omega}\,=\,
t_i(t_{1,\omega},q(t_{1,\omega}),\dots,t_{i-1,\omega},q(t_{i-1,\omega})),
\nonumber \\ 
\omega &\mapsto& \phi_\omega \nonumber,\\
\omega &\mapsto& n_{q,\omega} \nonumber
\end{eqnarray}
are measurable. Let $n_q=\E(n_{q,\omega})$ be the expected number of 
values of the function $q$ with respect to~$\omega$ . 
As before, we assume that 
$n\,=\,\sup_{q\in \Q}n_q<\infty$.  
The randomized error of such a randomized algorithm $\phi$ is given by
\begin{equation}
e^{\clr}(\phi, n)=\sup_{q\in \Q}\left( \E[\l(q) - \phi_\omega(q)]^2
\right)^{1/2}.
\label{eq:cre}
\end{equation}
For simplicity and brevity we consider the error of 
randomized algorithms in the $L_2$ sense.
It is straightforward to extend our results for the 
error of randomized algorithms defined in the $L_p$-sense 
with $p\in [1,\infty]$.

We denote the minimal number of function
values needed to compute an $\e$-approximation of the Sturm-Liouville 
eigenvalue problem in the worst case and randomized settings by
\begin{eqnarray*}
\nwor\,&=&\,\min\{\,n:\ \exists\ \phi \ \mbox{such that}\
e^{\cld}(\phi,n)\,\le\,\e\;\}\ \ \mbox{and}\\
\nran\,&=&\,\min\{\,n:\ \exists\ \phi \ \mbox{such that}\
e^{\clr}(\phi,n)\,\le\,\e\;\},
\end{eqnarray*}
respectively.

\subsection{Lower Bounds}

We now prove lower bounds on $\nwor$ and $\nran$.

\begin{thm}\label{thm1}
$$
\nwor\,=\,\Omega\left(\e^{-1/2}\right),\qquad
\nran\,=\,\Omega\left(\e^{-2/5}\right).
$$
\end{thm}
\vskip 1pc
\noindent {\it Proof.\ } Define
\begin{equation}\label{classF}
F\,=\,\left\{\,f:\ f \in C^2([0,1]),\
\max\left(\|f\|_{\infty},\|f^{\prime}\|_{\infty},
\|f^{\prime\prime}\|_{\infty} \left. \right)\,\le\,1\,\right\}\right),
\end{equation}
and consider the weighted integration problem
$$
I(f)\,=\int_0^1f(x)\sin^2(\pi x)\,dx\qquad \forall\,f\in F.
$$
It is well-known that any  algorithm using $n$ function values
for approximating of this weighted integration problem
has worst case error at least proportional to $n^{-2}$ in the
worst case setting, and to $n^{-2.5}$ in the randomized setting, see
\cite{novak,traub}\footnote{Formally, these results are proved for
$I(f)=\int_0^1f(x)\,dx$. However, the same proofs can be applied for the 
integration problem with the weight $\sin^2(\pi x)$ and the same lower
bounds hold.}. 

For $c\,\in\,(0,\tfrac12]$, consider the class
\begin{equation}\label{classfc}
F_c\,=\,F\,\cap\,\{\,f\in F\,: \ \|f\|_{\infty}\,\le\,c\,\}.
\end{equation}
For $n^{-2}$ much less than $c$, the proofs for the class $F$ can be
used to deduce the same lower bounds on algorithms for approximation
of the weighted integration problem for the class~$F_c$.

For $f\in F_c$ define $q=\tfrac12+f$. Then $q\in \Q$. {}From (\ref{444})
we have
$$
\l(q)\,=\,\pi^2\,+\,\tfrac12\,+\,2\,I(f)\,+\,O(c^2).
$$
For any algorithm $\phi$ using $n$ function values of $q$ for the
Sturm-Liouville eigenvalue problem, define the algorithm
$\psi(f)=\tfrac12(\phi(q)-\pi^2-\tfrac12)$ for the
weighted integration problem. Then $\psi$ uses $n$ function values of $f$, and
\begin{equation}\label{777}
\l(q)-\phi(q)\,=\,2\left(I(f)-\psi(f)\right)\,+\,O(c^2).
\end{equation}
Let $c=n^{-3/2}$. Then $n^{-2}=o(c)$, and therefore
the error of $\phi$ is lower bounded by $\Omega(n^{-2})$
in the worst case setting, and by $\Omega(n^{-2.5})$ in the
randomized setting. Hence, the error of $\phi$ is at most $\e$ when
$n=\Omega(\e^{-1/2})$ in the worst case setting, and
$n=\Omega(\e^{-2/5})$ in the randomized setting. Since this holds 
for an arbitrary algorithm $\phi$, the proof is complete. \qed

\subsection{Upper Bounds in the Worst Case Setting}

We now discuss upper bounds on $\nwor$, as well as 
bounds on the complexity in the worst case setting.
The worst case cost of an algorithm $\phi$
using $n$ function values is defined as 
$$
\costwor(\phi)\,=\,\sup_{q\in \Q}\left(\cc\, n_q + m_q\right) ,
$$
where $m_q$ is the number of arithmetic operations used by 
the algorithm for a function $q$ from~$\Q$. The worst case complexity
$\compwor(\e)$ is defined as the minimal cost of an algorithm whose
worst case error is at most $\e$,
$$
\compwor(\e)\,=\,\min\left\{\,\costwor(\phi)\,:\ \phi\ \mbox{such that}\
  e^{\cld}(\phi,n)\,\le\, \e\,\right\}.
$$
Obviously, $\compwor(\e)\,\ge\,\cc\,\nwor$.

We now discuss the classical algorithm for the Sturm-Liouville eigenvalue
problem, see e.g., \cite{demmel,keller}, and show that it is almost optimal
in the worst case setting. This algorithm uses $n=\Theta(\e^{-1/2})$
function values of $q$ at the equidistant points $i/(n+1)$ for
$i=1,2,\dots,n$. Then the operator $\L_q$ is approximated by the
tridiagonal $n\times n$ matrix $M_q$ of the form 
$$
M_q\,=\,(n+1)^2\,
\left[
\begin{array}{ccccc}
2 & -1 & & & \\
-1 & 2 & -1 & &\\
   &  \ddots & \ddots & \ddots & \\
   &         & -1 & 2 & -1 \\
   &         &    & -1 & 2
\end{array}
\right] + \left[
\begin{array}{ccccc}
q(\tfrac1{n+1}) & & & & \\
& q(\tfrac2{n+1}) & & &  \\
& & \ddots  & &  \\
& & & q(\tfrac{n-1}{n+1}) &  \\
& & &  & q(\tfrac{n}{n+1})
\end{array} \right].
$$

Clearly, $M_q$ is a symmetric and positive definite matrix. Let
$\l_j=\l_j(M_q)$ and $z_j=z_j(M_q)$ be the eigenvalues and eigenvectors of
$M_q$, i.e., $M_qz_j=\l_jz_j$ with
$$
\l_1\,\le\,\l_2\,\le\,\cdots\,\le\,\l_n,
$$
where the vectors $z_j$ are orthogonal and normalized such that
$$
\|z_j\|_{L_2}^2\,:=\,\frac1n\sum_{k=1}^nz_{j,k}^2=1
$$ 
with $z_{j,k}$ being the $k$th component of $z_j$. Note that we use
the subscript $L_2$ in the norm of a vector to stress 
similarity to the $L_2$ norm of functions, and to distinguish from
the Euclidean second norm. Clearly,
$\|z_j\|_{L_2}=\tfrac1{\sqrt{n}}\|z\|_2$. 

For $q\equiv c$, it is known, see, e.g., \cite{demmel}, that
$$
\l_j(M_c)\,=\,c\,+\,4(n+1)^2\sin^2\left(\frac{j\pi}{2(n+1)}\right),
$$
and $z_j(M_c)=[z_{j,1}(M_c),z_{j,2}(M_c),\dots,z_{j,n}(M_c)]^T$ with
$$
z_{j,k}(M_c)\,=\,\left(\frac{2n}{n+1}\right)^{1/2}\,
\sin\left(\frac{jk\pi}{n+1}\right).
$$

It is known, see, e.g., \cite{keller}, that the smallest eigenvalue
$\l_1(M_q)$ of the matrix $M_q$ approximates the smallest eigenvalue $\l(q)$ of
the operator $\L_q$ with error of order $n^{-2}$, i.e.,
$$
\l(q)\,-\,\l_1(M_q)\,=\,O\left(n^{-2}\right)\,=\,O(\e).
$$
Hence, it is enough to approximate $\l_1(M_q)$ with error of order
$\e$. This can be achieved by using roughly $\log\,\e^{-1}$ bisection
steps. Each step consists of computing the $n$ terms of the Sturm
sequence, and this can be done in cost proportional to $n$.
The total cost is of order $(\cc\,+\log\,\e^{-1})\e^{-1/2}$. For
details, see \cite{demmel,wilkinson}. 
Theorem \ref{thm1} and the cost of this algorithm lead to 
the following bounds for the minimal number of function values
and for the worst case complexity. 
\begin{thm}
$$
\nwor\,=\,\Theta(\e^{-1/2}),\qquad
\Omega(\cc\,\e^{-1/2})= \compwor(\e)\,=\, O(\cc\,\e^{-1/2}\,+\,
\e^{-1/2}\log\,\e^{-1}).
$$
\end{thm}
\vskip 2pc
{\bf Remark 4.1.} \
We now show how (\ref{222}) can be proven, based on the properties of
the matrix~$M_q$. First observe that for $q=0$, the eigenvalues
$\l_j(M_0)$ are well separated, since
\begin{eqnarray*}
\l_{j+1}(M_0)-\l_j(M_0)\,&=&\,4(n+1)^2\sin\frac{(2j+1)\pi}{2(n+1)}\,
\sin\frac{\pi}{2(n+1)}\\
\,&\ge&\,4(n+1)^2\sin\frac{3\pi}{2(n+1)}\,
\sin\frac{\pi}{2(n+1)}\,\approx\, 3\pi^2. 
\end{eqnarray*}
For $q\in\Q$, the Hermitian matrix $M_q$ differs from $M_0$ by the
diagonal matrix diag$\,q(i/(n+1))$ whose elements satisfy
$q(i/(n+1))\in[0,\|q\|_{\infty}]$ with $\|q\|_{\infty}\le1$. Using the
known estimates on the perturbed eigenvalues of Hermitian matrices, 
see \cite{wilkinson}, we have 
$$
\min_{i=1,2,\dots,n}\left|\l_j(M_q)-\l_i(M_0)\right|\,\le\,\|q\|_{\infty}
$$
for all $j=1,2,\dots,n$. Since the intervals
$[\l_i(M_0)-1,\l_i(M_0)+1]$ are disjoint, we conclude that
$$
\left|\l_j(M_q)-\l_j(M_0)\right|\,\le\,\|q\|_{\infty}\,\le\,1,
$$
and that
$$
\l_{j+1}(M_q)-\l_j(M_q)\,\ge\,\l_{j+1}(M_0)-\l_j(M_0)-2
\,\approx\, 3\pi^2-2. 
$$
Define
$$
\tilde{u}_{q,n}\,=\,\left[u_q\left(\frac1{n+1}\right),\dots,
u_q\left(\frac{n}{n+1}\right)\right]^T,
$$
where $u_q$ is the normalized real eigenfunction corresponding to the smallest
eigenvalue.
Then $\|\tilde{u}_{q,n}\|_{L_2}=1+o(1)$. 
We normalize $\tilde{u}_{q,n}$ and obtain
$$
u_{q,n}\,=\,\frac1{\|\tilde{u}_{q,n}\|_{L_2}}\,\tilde{u}_{q,n}.
$$

As mentioned in Section 3, the eigenfunction $u_q$ is defined
uniquely up to its sign. Obviously, the same is true for the\
eigenvector $z_1(M_q)$. We choose the signs of $u_q$ and $z_1(M_q)$
such that
$$
\|u_{q,n}-z_1(M_q)\|_{L_2}\,\le\,\|u_{q,n}+z_1(M_q)\|_{L_2}.
$$
All the components of the vector
$$
\eta_n\,:=\,M_qu_{q,n}-\l(q)u_{q,n}
$$
are of order $n^{-2}$, and therefore
$\|\eta_n\|_{L_2}=O(n^{-2})$. {}From the a posteriori error estimate,
see \cite[p.~173]{wilkinson}, we conclude that
$$
\|u_{q,n}-z_1(M_q)\|_{L_2}\,=\,O(n^{-2})\qquad \forall\, q\in \Q
$$
with the factor in the big-$O$ notation independent of $q$. Note also
that 
$$
M_qu_{\bar q,n}-\l(q)u_{\bar q,n}\,=\,M_{\bar q}u_{\bar q,n}-\l(\bar
q)u_{\bar q,n}\,+r_n,
$$
with $\|r_n\|_{L_2}=O(\|q-\bar q\|_{\infty})$. Hence
$$
\|u_{\bar q,n}-z_1(M_q)\|_{L_2}\,=\,O(\|q-\bar q\|_{\infty}+n^{-2}).
$$
Finally, we have
$$
\|u_{q,n}-u_{\bar q,n}\|_{L_2}\,=\,\|u_{q,n}-z_1(M_q)+z_1(M_q)-u_{\bar
q,n}\|_{L_2} \,=\, O(n^{-2}+\|q-\bar q\|_{\infty}).
$$
Letting $n$ tend to infinity, we conclude that
$$
\|u_q-u_{\bar q}\|_{L_2}\,=O(\|q-\bar q\|_{\infty}).
$$
Since both $u_q$ and $u_{\bar q}$ satisfy (\ref{eq:SL1}) for
$(q,\l(q))$ and $(\bar q,\l(\bar q))$, respectively, we have
$$
u^{\prime\prime}_q(x)-u^{\prime\prime}_{\bar q}(x)\,=\,
(q(x)-\l(q))(u_q(x)-u_{\bar q}(x))\,+\,u_{\bar q}(x)\left(
(q(x)-\bar q(x))-(\l(q)-\l(\bar q))\right).
$$
Therefore
$$
\|u^{\prime\prime}_q-u^{\prime\prime}_{\bar q}\|_{L_2}\,=\,
O(\|q-\bar q\|_{\infty}).
$$
This and the fact that $u-u_{\bar q}$ vanishes at $0$ and $1$ imply 
$$
\|u_q-u_{\bar q}\|_{\infty}\,=O(\|q-\bar q\|_{\infty}),
$$
as claimed.  \qed

\subsection{Upper Bounds in the Randomized Setting}

We now turn to the randomized setting. 
The cost of a randomized algorithm $\phi$, using
$n=\sup_{q\in\Q}\E(n_{q,\omega})<\infty$ 
randomized function values, is now defined as
$$
\costran(\phi)\,=\,\sup_{q\in \Q}\left(\E\left(\cc\, n_{q,\omega} +
 m_{q,\omega}\right)^2\right)^{1/2} ,
$$
where $m_{q,\omega}$ is the number of arithmetic operations used by 
the algorithm for a function $q$ from~$\Q$ and a random variable
$\omega$. The randomized complexity
$$
\compran(\e)\,=\,\min\left\{\,\costran(\phi)\,:\ \phi\ \ 
   \mbox{such that}\ \ 
  e^{\clr}(\phi,n)\,\le\, \e\,\right\},
$$
is the minimal cost of an algorithm whose randomized error is at most $\e$.
Obviously, $\compran(\e)\,\ge\,\cc\,\nran$.

We now derive upper bounds on $\nran$ and $\compran(\e)$ by presenting
a randomized algorithm that depends on a number of parameters. 
Then we find the values of these parameters for which
the randomized error is $\e$. 
We first compute $m+1$ function values of $q$ at deterministic points
$i/m$, for $i=0,1,\dots,m$, 
and construct a cubic natural spline
$q_{{\rm cub}}$ interpolating $q$ at these points, see
e.g., \cite{CK} for information about cubic splines. 
It is well known that
this can be done with cost proportional to~$m$, and
$\|q-q_{{\rm cub}}\|_{\infty}=O(m^{-2})$.
The function $q_{{\rm cub}}$ does not have to be non-negative.
Since $q\ge0$ then $\bar q = q_{{\rm cub}}+c\ge0$ with a constant
$c=O(m^{-2})$. We have $\bar q\in \Q$ and $\|q-\bar
q\|_{\infty}=O(m^{-2})$. We apply the formula (\ref{333}) for the
function $\bar q$ and obtain 
\begin{equation}\label{999}
\l(q)\,-\,\l(\bar q)\,=\,\int_0^1\left(q(x)-\bar q(x)\right)
u_{\bar q}^2(x)\,dx
\,+\, O\left(m^{-4}\right).
\end{equation}
This suggests that we can improve the accuracy of approximating 
$\l(q)-\l(\bar q)$ by using the classical Monte Carlo algorithm  
applied to the first term of the right hand side of 
(\ref{999}). We will need to know,
at least approximately, the eigenvalue $\l(\bar q)$ and the
eigenvector $u_{\bar q}$.  Suppose we approximate $\l(\bar q)$ by
$\l_{\bar q}$ with the worst case error
\begin{equation}\label{1111}
\sup_{q\in \Q}\left|\l(\bar q)-\l_{\bar q}\right|\,\le\, \delta_1,
\end{equation}
and the eigenfunction $u_{\bar q}$ by $z_{\bar q}$ with the worst case error 
\begin{equation}\label{2222}
\sup_{q\in \Q}\|u_{\bar q}-z_{\bar q}\|_{L_2}\,\le\,\delta_2.
\end{equation}

Assume for a moment that $\l_{\bar q}$ and $z_{\bar q}$ have
been computed. For a function $v$, define $f_v(x)=(q(x)-\bar q(x))v^2(x)$ and 
$I(f_v)=\int_0^1f_v(x)\,dx$. 

The randomized algorithm $\phi$ based on the Monte Carlo with 
$k$ randomized samples takes the form
$$
\phi_{\omega}(q)\,=\,\l_{\bar q}\,+\,\frac1k\sum_{j=1}^k\bigg(q(x_{j,\omega})
\,-\,\bar q(x_{j,\omega})\bigg)z_{\bar q}^2(x_{j,\omega}),
$$
where $x_{j,\omega}$ are independent and uniformly distributed
numbers from $[0,1]$. Here $\omega$ represents a random element. We have
\begin{eqnarray*}
\left|\l(q)-\phi_{\omega}(q)\right|\,&\le&\,|\l(\bar q)-\l_{\bar q}|\,+\,
|I(f_{u_{\bar q}})-I(f_{z_{\bar q}})|\\
&+&\,
\bigg|I(f_{z_{\bar q}})-\frac1k\sum_{j=1}^kf_{z_{\bar q}}
(x_{j,\omega})\bigg|\,+\,O(m^{-4}).
\end{eqnarray*}
Clearly,
$$
\left|I(f_{u_{\bar q}})-I(f_{z_{\bar q}})\right|\,\le\,
\int_0^1\left|q(x)-\bar q(x)\right|\left|u_{\bar
q}^2(x)-z_{\bar q}^2(x)\right|\,dx\,=\,
O(m^{-2}\,\delta_2).
$$
Since $\|f_{z_{\bar q}}\|_{L_2}=O(m^{-2})$, 
the well known formula for the randomized error of Monte Carlo
yields that
$$
\left(\E_{\omega}\left(I(f_{z_{\bar q}}) -\frac1k\sum_{j=1}^k
f_{z_{\bar q}}(x_{j,\omega})\right)^2\right)^{1/2}\,=\,
\frac{(I(f^2_{z_{\bar q}})-I^2(f_{z_{\bar q}}))^{1/2}}{k^{1/2}}\,=\,
O\left(m^{-2}k^{-1/2}\right).
$$

We have obtained the bound 
$$
e^{\clr}(\phi,n)\,=\,O\left(\delta_1+m^{-2}\delta_2+m^{-2}k^{-1/2}+m^{-4}
\right)
$$
on the randomized error of $\phi$.
Hence, to guarantee error at most $\e$, it is enough to take 
$$
\delta_1=\Theta(\e),\ \ m=k=\Theta(\e^{-2/5})\ \
\mbox{and}\ \  \delta_2=\Theta(\e^{1/5}).
$$

We now explain how to achieve (\ref{1111}) and (\ref{2222}). 
To get $\l_{\bar q}$ approximating $\l(\bar q)$ with error of order $\e$, we 
approximate the operator $\L_{\bar q}$ by the matrix $M_{\bar q}$ as in the
worst  case setting, now with $n=\Theta(\e^{-1/2})$. Then 
$\l({\bar q})-\l_1(M_{\bar q})=O(n^{-2})=O(\e)$, and we compute
$\l_{\bar q}$ as an $\e$-approximation of $\l_1(M_{\bar q})$ as for
the worst case setting. This can be done with cost of order $\e^{-1/2}$
function values of $\bar q$, and of order $\e^{-1/2}\log\,\e^{-1}$
arithmetic operations. Since the cost of computing one function value
of $\bar q$ is of order $1$, the total cost of computing $\l_{\bar q}$ is of
order $\e^{-1/2}\log\,\e^{-1}$. 

To get $z_{\bar q}$ approximating $u_{\bar q}$ with
error of order $\e^{1/5}$ we proceed as follows. 
Consider the eigenvector $z_1(M_{\bar q})$ of the matrix $M_{\bar q}$,
with $n$ not yet specified. By Remark 4.1, we have
\begin{equation}\label{3333}
\|u_{\bar q,n} - z_1(M_{\bar q})\|_{L_2}\,=\,O(n^{-2}).
\end{equation}
We approximate the smallest eigenvalue $\l_1(M_{\bar q})$ by $\bar \l$,
with error $\delta$. This can be achieved with cost of order
$n\log\,\delta^{-1}$. Without loss of generality we assume that
$\bar \l\not=\l_1(M_{\bar q})$. Indeed, we can check this condition by
computing $\mbox{det}(M_{\bar q}-{\bar \l}I)$ and if this determinant
is zero we perturb $\bar \l$ a little. Then the matrix
$$
A\,=\,\left(M_{\bar q}-{\bar \l}I\right)^{-1}
$$
is non-singular and its eigenvalues are 
$\beta_j=(\l_j(M_{\bar q})-{\bar \l})^{-1}$. Note that $|\beta_1|\ge
\delta^{-1}$ and $\beta_j=O(1)$ for $j\ge2$. For the $j$th vector 
$e_j=[0,\dots,0,1,0,\dots,0]^T$ with $1$ in the $j$th position, define
$$
x_j\,=\,A\,e_j.
$$
We can compute $x_j$ with cost of order $n$ by solving the tridiagonal linear
system $(M_{\bar q}-{\bar \l}I)x_j=e_j$. Then we compute 
$$
\|x_{j_0}\|_2\,=\,\max_{j=1,2,\dots,n}\|x_j\|_2,
$$
and 
$$
z\,=\,\|x_{j_0}\|^{-1}_2\,x_{j_0}.
$$
Observe that the cost of computing $z$ is of order $n^2$.  

Since $\{n^{-1/2}z_j(M_{\bar q})\}_{j=1}^n$ is orthonormal, we have 
$\|x_j\|^2_2=\sum_{\ell=1}^n\beta^2_{\ell}(e_j,n^{-1/2}\,z_{\ell}(M_{\bar
q}))^2$ and $\|n^{-1/2}\,z_1(M_{\bar q})\|^2_2=1=\sum_{j=1}^n
(e_j,n^{-1/2}\,z_1(M_{\bar q}))^2$. Hence, there exists an index $j$ such
that 
$$
(e_j,n^{-1/2}\,z_1(M_{\bar q}))^2\,\ge\, n^{-1},
$$
and therefore
$$
\|x_{j_0}\|_2\,\ge\,\|x_j\|_2\,\ge\,\delta^{-1}n^{-1/2}.
$$
We have
$$
(M_{\bar q}-\l_1(M_{\bar q})I)z\,=\, (M_{\bar q}-{\bar \l}I)z\,+\,
(\bar \l -\l_1(M_{\bar q}))z\,=\,\frac1{\|x_{j_0}\|_2}e_{j_0}\,+\,
(\bar \l -\l_1(M_{\bar q}))z,
$$
and therefore
$$
\|(M_{\bar q}-\l_1(M_{\bar q})I)z\|_2\,\le\,\delta \sqrt{n}+\delta.
$$
{}From \cite[p.~173]{wilkinson}, we conclude that
$\|n^{-1/2}\,z_1(M_{\bar q})-z\|_{2}\,=\,O(\delta\,\sqrt{n})$, and 
\begin{equation}\label{5555}
\|z_1(M_{\bar q})-\sqrt{n}\,z\|_{L_2}\,=\,O(\delta \sqrt{n}).
\end{equation}

We are finally ready to define $z_{\bar q}$ by piecewise linear
interpolation from the successive components of the vector
$\sqrt{n}\,z=[z_1,z_2,\dots,z_n]^T$. More precisely,
for $j=0,1,\dots,n$ let $t_j=j/(n+1)$. For $t\in[t_j,t_{j+1}]$, 
we set
$$
z_{\bar q}(t)\,=\,z_j (1-(n+1)t+j)\,+\,z_{j+1} ((n+1)t-j)
$$
with $z_0=z_{n+1}=0$. 

We need to estimate $u_{\bar q}-z_{\bar q}$ in the
$L_2$ norm. Observe that for $t\in[t_j,t_{j+1}]$ we have
$$
u_{\bar q}(t)\,=\,u_{\bar q}(t_j)(1-(n+1)t+j)\,+\,u_{\bar
  q}(t_{j+1})((n+1)t-j)\,+\,O(n^{-2})
$$
since $u_{\bar q}\in \Q$. Therefore
$$
|u_{\bar q}(t)-z_{\bar q}(t)|\,\le\,
|u_{\bar q}(t_j)-z_{\bar q}(t_j)|\,+\,
|u_{\bar q}(t_{j+1})-z_{\bar q}(t_{j+1})|\,+\,O(n^{-2}). 
$$
This yields
\begin{eqnarray*}
\|u_{\bar q}-z_{\bar
  q}\|^2_{L_2}\,&=&\,\sum_{j=0}^n\int_{t_j}^{t_{j+1}}
\left(u_{\bar q}(t)-z_{\bar q}(t)\right)^2dt\\
&=&\,O\left(\frac1{n+1}\sum_{j=0}^n\left(u_{\bar q}(t_j)-
z_{\bar q}(t_j)\right)^2\,+\,n^{-4}\right).
\end{eqnarray*}
Hence,
$$
\|u_{\bar q}-z_{\bar q}\|_{L_2}\,=\,O\left(\|u_{\bar
  q,n}-\sqrt{n}\,z\|_{L_2}\,+n^{-2}\right)
$$
Since $\|u_{\bar q,n}-\sqrt{n}\,z\|_{L_2}\,\le
\|u_{\bar q,n}-z_1(M_{\bar q})\|_{L_2}\,+\,
\|z_1(M_{\bar q})-\sqrt{n}\,z\|_{L_2}$, we use
(\ref{3333}) and (\ref{5555}) to see that
$$
\|u_{\bar q}-z_{\bar q}\|_{L_2}\,=\,O(\delta \sqrt{n}+n^{-2}).
$$
For $\delta=n^{-5/2}$ we obtain
$$
\|u_{\bar q}-z_{\bar q}\|_{L_2}\,=\,O(n^{-2}).
$$
Setting $n=\Theta(\e^{-1/10})$ we obtain (\ref{2222}) with
$\delta_2=\Theta(\e^{1/5})$. The cost of computing $z_{\bar q}$ is of
order $n^2=\Theta(\e^{-1/5})$.

Theorem \ref{thm1} and the cost of this randomized algorithm lead to
the following bounds on the minimal number of function values and the
randomized complexity.

\vskip 1pc
\begin{thm}
$$
\nran\,=\,\Theta(\e^{-2/5}),\qquad
\Omega(\cc\,\e^{-2/5})=  \compran(\e)\,=\, O(\cc\,\e^{-2/5}\,+\,
\e^{-1/2}\log\,\e^{-1}).
$$
\end{thm}

\section{Quantum Setting}

We now turn our attention to the quantum setting.
In this setting, we are using {\it hybrid} algorithms that are
combinations of classical algorithms using function values, as
explained in the previous sections, and quantum algorithms which we now 
describe. A quantum algorithm applies a sequence of unitary 
transformations to an initial state, and 
the final state is measured, see \cite{beals,cleve,heinrich,nielsen} 
for the details of the quantum model of computation. We briefly 
summarize this model to the extent necessary for this paper.

The initial state $|\psi_0\rangle$ is a unit vector of the Hilbert space
$\Cal{H}_\nu=\complex^2\otimes \cdots\otimes \complex^2$, $\nu$ times,  
for some appropriately chosen integer $\nu$, where $\complex^2$ is the 
two dimensional space of complex numbers. Obviously, the dimension of
$\Cal{H}_\nu$ is $2^{\nu}$. The number $\nu$ denotes the number of 
qubits used in quantum computation. 

The final state $|\psi\rangle$ is also a unit vector of
$\Cal{H}_\nu$ and is obtained from the initial state 
$|\psi_0\rangle$ by
applying a number of unitary $2^{\nu}\times 2^{\nu}$ matrices, i.e.,
\begin{equation}
|\psi\rangle\,:=\,U_TQ_YU_{T-1}Q_Y\cdots U_1Q_YU_0 |\psi_0\rangle.
\label{eq:qa}
\end{equation}
Here, $U_0,U_1,\dots,U_T$ are unitary matrices that do not depend 
on the input function $q$. The unitary matrix $Q_Y$ with 
$Y=[q(t_1),\dots,q(t_n)]$ is called a quantum query and depends on
$n$, with $n\le 2^{\nu}$, 
function evaluations of $q$ computed at some non-adaptive points
$t_i\in[0,1]$. The quantum query $Q_Y$ is the only source of
information about $q$. The integer $T$ denotes the number of quantum
queries we choose to use.

At the end of the quantum algorithm, a measurement is 
applied to its final state $|\psi \rangle$.
The measurement produces  one of $M$ outcomes, where $M\le 2^{\nu}$. 
Outcome $j\in\{0,1,\dots,M-1\}$ occurs with
probability $p_Y(j)$, which depends on $j$ and the input $Y$.
For example, if $M=2^{\nu}$ and the final state is
$|\psi\rangle=\sum_{j=0}^{2^\nu-1}c_j|j\rangle$,
with $\sum_{j=0}^{2^\nu-1}|c_j|^2=1$, then a measurement 
in the computational orthonormal basis $\{|j\rangle\}$ produces 
the outcome $j$ with probability $p_Y(j)=|c_j|^2$. 
Knowing the outcome $j$, we compute
an approximation $\hat\l_Y(j)$ of the smallest eigenvalue on a
classical computer.  

In principle, quantum algorithms may have many measurements
applied between sequences of unitary transformations of the form 
presented above. However, any algorithm with many measurements and 
a total of $T$ quantum queries can be simulated by
a quantum algorithm with only one measurement at the end,
for details see e.g., \cite{heinrich}. 

We stress that classical algorithms in floating or fixed point
arithmetic can also be written in the form of 
(\ref{eq:qa}). Indeed, all classical bit operations 
can be simulated by quantum computations, see e.g., \cite{bernstein}. 
Classically computed function values will correspond to bit queries
which we discuss in Section 5.2. 

In our case, we formally use the real number model of computation. Since
the Sturm-Liouville eigenvalue problem is well conditioned and 
properly normalized, we obtain practically the same results in floating
or fixed point arithmetic. More precisely, it is enough to use
$O(\log\,\e^{-1})$ mantissa bits, and the cost of bit operations 
in floating or fixed point arithmetic is of the same order as the cost
in the real number model multiplied by a power of $\log\,\e^{-1}$. 

Hence, a hybrid algorithm may be viewed as a finite sequence of algorithms 
of the form (\ref{eq:qa}). It is also known that if we use 
finitely many algorithms of
the form (\ref{eq:qa}) then they can be written as one quantum
algorithm of the form (\ref{eq:qa}), see \cite{heinrich,H03}. 

That is why an arbitrary hybrid algorithm in the quantum setting 
is of the form (\ref{eq:qa}). This is important when we want to
prove lower bounds because it is enough to work with 
algorithms of the form (\ref{eq:qa}). For upper bounds, it seems to us
more natural to distinguish between classical and quantum computations and
charge their cost differently.
The cost of classical computations is defined as before whereas 
the cost of quantum computations is defined as the sum of  the number
of quantum queries multiplied by the cost of one query, and  
the number of quantum operations besides quantum queries.
It will be also important to indicate how many qubits are used by
the quantum computations.

We now define the error in the quantum setting.
In this setting,  we want to approximate the smallest eigenvalue
$\l(q)$ with a probability $p>\tfrac12$. For simplicity, we take
$p=\tfrac34$ for the rest of this section. 
As it is common for quantum algorithms, we can achieve an
$\e$-approximation with probability arbitrarily close to
$1$ by repetition of the original quantum algorithm, and by taking
the median as the final approximation. 

The local error of the quantum
algorithm with $T$ queries that computes $\hat \l_Y(j)$ for the function
$q\in \Q$ and the outcome $j\in\{0,1,\dots,M-1\}$ is defined by 
\begin{equation*}
e(\hat\l_Y,T)\,=\,\min \bigg\{\, \a :\quad \sum_{j:\ |\l(q) - \hat
\l_{Y}(j)|\,\le\, \a\,}p_Y(j)\geq \tfrac34\,\bigg\}.
\label{eq:localperr}
\end{equation*}
This can be equivalently rewritten as
$$
e(\hat\l_Y, T)\,=\,\min_{A:\, \mu(A)\ge \tfrac34}\max_{j\in A}
\big|\l(q)-\hat \l_Y(j)\big|,
$$
where $A\subset\{0,1,\dots,M-1\}$ and $\mu(A)=\sum_{j\in A}p_Y(j)$. 

The {\it worst probabilistic} error of a quantum algorithm $\hat\l$ with $T$ 
queries for the Sturm-Liouville eigenvalue problem is defined by
\begin{equation}
e^{\w}(\hat\l, T)\,=\,\sup\bigg\{\, 
e(\hat \l_Y,T)\colon \ Y=[q(t_1),\dots,q(t_n)],\ \ t_i\in [0,1], \ \
\mbox{for}\ \ q\in \Q \,\bigg\}.
\label{eq:wperr}
\end{equation}

\subsection{Bit Queries}

Quantum queries are important
in the complexity analysis of quantum algorithms.
A quantum query corresponds to a function 
evaluation in classical computation. 
By analogy with the complexity analysis of classical algorithms,
we analyze the cost of quantum algorithms in
terms of the number of quantum queries that are necessary to compute an
$\e$-approximation with probability~$\tfrac34$. 
Clearly, this number is a lower bound on the quantum complexity, which
is defined as the minimal total cost of a quantum
algorithm that solves the problem. 

Different quantum queries have been studied in the literature.
Probably the most commonly studied query is the {\it bit} query.
For a Boolean function $f:\{0,1,\dots,2^m-1\}\to\{0,1\}$, 
the bit query is defined by  
$$
Q_f|j\rangle|k\rangle\,=\,|j\rangle|k\oplus f(j)\rangle. 
$$
Here $\nu=m+1$, $|j\rangle\in\Cal{H}_m$, and
$|k\rangle\in\Cal{H}_{1}$ with $\oplus$ denoting the addition
modulo $2$. For real functions, such as functions $q$, the bit query
is constructed by taking the 
most significant bits of the function $q$ evaluated at some points
$t_j$.
More precisely, as in \cite{heinrich}, the bit query for $q$ has the
form
$$
Q_q|j\rangle|k\rangle\,=\,|j\rangle|k\oplus \beta(q(\tau(j)))\rangle, 
$$
where the number of qubits is now $\nu=m'+m''$ and $|j\rangle\in
\Cal{H}_{m'}$, $|k\rangle\in\Cal{H}_{m''}$ with some functions
$\beta:[0,1]\to\{0,1,\dots,2^{m''}-1\}$ and
$\tau:\{0,1,\dots,2^{m'}-1\}\to[0,1]$. Hence, we compute $q$ at 
$t_j=\tau(j)\in[0,1]$ and then take the $m''$ most significant bits of
$q(t_j)$ by $\beta(q(t_j))$, for details and a possible use
of ancilla qubits see again \cite{heinrich}.

Using bit queries, the well known
quantum algorithm of Grover \cite{grover} 
requires $\Theta(N^{1/2})$ queries for searching an unordered
database of $N$ items. 
Similarly, the quantum summation 
algorithm of Brassard et al.~\cite{brassard} 
computes the mean of a Boolean function defined on 
the set of $N$ elements with 
accuracy $\e$ and probability $\tfrac34$ using of order 
$\min\{N, \e^{-1}\}$ bit queries.
Both algorithms are optimal modulo multiplicative factors
in terms of the number of bit queries.

The quantum summation algorithm can be also used for the approximate
computation of the mean of a real function 
$f:[0,1]\to\reals$ with $|f(x)|\le M$ for all 
$x\in[0,1]$, see \cite{heinrich,novak}. More precisely, if we want to
approximate
$$
\mbox{S}_N(f)\,:=\,\frac1N\sum_{j=0}^{N-1}f(x_j)
$$
for some $x_j\in [0,1]$ and $N$, then the quantum summation algorithm
$\mbox{QS}_N(f)$ approximates $\mbox{S}_N(f)$ such that
\begin{equation}\label{157}
|\mbox{S}_N(f)-\mbox{QS}_N(f)|\,\le\,\e \qquad\mbox{with probability}\ \tfrac34
\end{equation}
using of order $\min(N,M\e^{-1})$ bit queries,
$\min(N,M\e^{-1})\,\log\,N$ quantum operations, and
$\log\,N$ qubits. 

Bit queries have been also used
for a number of continuous problems such as multivariate and path
integration, multivariate approximation, and ordinary differential
equations. Tight bit query complexity
bounds are known for a number of such problems, see
\cite{heinrich,H03,H04a,H04b,Kacewicz,N01,TW02}.  

In particular, Novak \cite{N01} proved that for the integration
problem $\int_0^1f(x)\,dx$ for functions $f$ from the class $F$ 
given by (\ref{classF}), the bit query complexity is 
\begin{equation}\label{novak}
n^{{\rm bit-query}}(\e,\mbox{INT}_F)\,=\,\Theta(\e^{-1/3}).
\end{equation} 
Here and elsewhere by the bit query complexity we understand the
minimal number of bit queries needed to compute an
$\e$-approximation to a given problem with probability~$\tfrac34$. 
In particular, 
$n^{\textrm{\rm bit-query}}(\e)$ 
denotes the bit query complexity of
the Sturm-Liouville eigenvalue problem.

Based on the result (\ref{novak}) of Novak and the relationship 
between the Sturm-Liouville eigenvalue problem with integration, 
we now prove the following theorem. 
\begin{thm}
\begin{equation*}
n^{\textrm{\rm bit-query}}(\e)\,=\,\Omega(\e^{-1/3}). 
\end{equation*}
\end{thm} 
\vskip 1pc
\noindent {\it Proof.\ } We first prove that the bit query complexity
for the weighted integration problem for the class $F_c$ given by
(\ref{classfc}) is of the same order as for integration for the class $F$,
\begin{equation}\label{6666}
n^{\textrm{\rm bit-query}}(\e,\mbox{INT}_{F_c})\,=\,\Theta(\e^{-1/3}).
\end{equation}
The upper bound follows from (\ref{novak}). To prove the lower
bound, we use the standard proof technique of reducing
the integration problem to the mean Boolean summation problem for which
a lower bound on bit queries is known. 

Assume then that we use an arbitrary quantum algorithm with $k$ 
bit queries that computes an $\e$-approximation with probability
$\tfrac34$ for the integration problem over the class $F_c$. Without
loss of generality we assume that $k^{-2}\le c$.  

Consider the function $h(x)=\a x^3(1-x)^3$ for $x\in [0,1]$ and
$h(x)=0$ for $x>1$. Here, $\a$ is a positive number chosen such that
$h\in F$ with $F$ given by (\ref{classF}).
For $j=0,1,\dots,N-1$, with $N>k$, define $h_j(x)=N^{-2}h(N(x-j/N))$. 
Clearly, $h_j\in F$ and the support of $h_j$ is $(j/N,(j+1)/N$. 
Observe that $\|h_j\|_{\infty}\le N^{-2}$. Hence $h_j\in F_c$. We also
have $\int_0^1h_j(x)\,dx=N^{-3}\int_0^1h(x)\,dx$. 
For an arbitrary Boolean function $B:\{0,1,\dots,N-1\}\to\{0,1\}$,
define the function
$$
f_B(x)\,=\,\sum_{j=0}^{N-1}B(j)h_j(x)\quad \forall\,x\in [0,1].
$$
Then $f_B\in F_c$ and
$$
\int_0^1f_B(x)\,dx\,=\, \frac{\int_0^1h(x)\,dx}{N^2}\ \frac
1N\sum_{j=0}^{N-1}B(j).  
$$
Hence, modulo the factor of order $N^{-2}$, the 
computation of the Boolean mean is reduced to the integration problem.
Note that $f_B(t)=B(j)h_j(t)$ if $t\in [j/N,(j+1)/N]$, 
and sampling of $f_B$ is equivalent to sampling of $B$.
{}From \cite{nayak} we know that $\Omega(k^{-1})$ is a lower bound 
for the error of the quantum approximation of the Boolean mean,
with $k$ bit queries, and probability $\tfrac34$, 
where $N\ge \beta k$ for some positive $\beta$.
Letting $N=\lceil \beta k \rceil$, we conclude that 
the corresponding lower bound 
on the integration problem over the class $F_c$
is $\Omega(k^{-3})$.  Hence to achieve the error $\e$ we
must have $k=\Omega(\e^{-1/3})$, as claimed in (\ref{6666}).

The same proof techniques allows us to consider the classes
$F_{c(\e)}$ with varying $c(\e)$, even with $c(\e)$ tending to zero,
although not too fast. We have
\begin{equation}\label{ctozero}
n^{\text{bit-query}}(\e,\mbox{INT}_{F_{c(\e)}})\,=\,\Theta(\e^{-1/3})
\qquad \mbox{if}\ \lim_{\e\to 0}c(\e)\,\e^{-2/3}\,=\,\infty.
\end{equation}

We now turn to the Sturm-Liouville eigenvalue problem. 
As in the proof of Theorem~\ref{thm1}, for $f\in F_c$ with
$c\in(0,\tfrac12]$, we define
$q=\tfrac12+f$ and consider an arbitrary quantum algorithm $\phi$ that
uses $k$ quantum bit queries and computes an $\e$-approximation of the
smallest eigenvalue with probability $\tfrac34$. Then
$\psi(f)=\tfrac12(\phi(q)-\pi^2-\tfrac12)$ is a quantum algorithm for
approximating the integration problem over the class $F_c$. We have 
$$
\left|I(f)-\psi(f)\right|\,=\,\left|\tfrac12\left(\l(q)-
\phi(q)\right)\,+\,O(c^2)\right|\,\le\,
\tfrac12\,\e+O(c^2).
$$
Take now $c=c(\e)=\Theta(\e^{2/3-\delta})$ with
$\delta\in(0,\tfrac16)$. Then
$$
\left|I(f)-\psi(f)\right|\,\le\,
\tfrac12\,\e+O(\e^{4/3-2\delta})\,=\,\tfrac12\,\e(1+o(1))\,
\le\,\e\quad\mbox{for small}\ \e.
$$
Hence, the quantum error of $\psi$ with
probability $\tfrac34$ is $\e$, and $\psi$ uses $k$ bit queries.
Due to (\ref{ctozero}), we have $k=\Omega(\e^{-1/3})$ which completes the
proof.  \  \ \qed

\vskip 1pc
We now derive upper bounds on the bit query complexity 
$n^{\text{bit-query}}(\e)$ and on the total quantum complexity 
${\rm comp}^{\text{bit-quant}}(\e)$. The total quantum complexity
is defined as the minimal cost of a hybrid algorithm that solves
the Sturm-Liouville eigenvalue problem with error at most $\e$ and
probability $\tfrac34$. The hybrid algorithm may require some 
classical computations and the use of function values and the cost
of them is defined just as before. It may also require 
some quantum computations and the cost
of them is defined as the sum of the number of bit queries multiplied by 
the cost of one such query plus the number of additional quantum operations.
The cost of one bit query is denoted by $\cc_{{\rm bit}}$.

We present a hybrid algorithm,
which will be a combination of the classical algorithm from Section 4
and the quantum summation algorithm $\mbox{QS}_N$ for a properly chosen $N$.
We proceed as in Section 4 and use the same notation. {}From
(\ref{999}), (\ref{1111}), and (\ref{2222}), we have
\begin{equation}\label{bit1}
\l(q)\,=\,\l_{\bar q}\,+\,\int_0^1(q(x)-\bar q(x))z_{\bar q}(x)\,dx 
\,+\,O(\delta_1+m^{-2}\delta_2+m^{-4})
\end{equation}
with $\delta_1,\,\delta_2$ and $m$ to be specified later. Let
$$
f(x)\,=\,(q(x)-\bar q(x))z_{\bar q}(x)\qquad x\in[0,1].
$$
Observe that $f(x)=O(m^{-2})$, and $f(x)$ depends on $q(x)$, and
$q(i/m)$ for $i=0,1,\dots,m$, which are used in the construction of
$\bar q$. 
Furthermore, we can compute $f(x)$
by computing one function value $q(x)$ and one function value of the
already computed functions $\bar q$ and $z_{\bar q}$ at $x$. 
We approximate $\int_0^1f(x)\,dx$ by
$$
\mbox{S}_N(f)\,=\,\frac1N\sum_{j=0}^{N-1}f\left(\frac{j+1}N\right)
$$
with $N=(m+1)k$, where the parameters $m$ and $k$ will be specified later. 
Since $f$ is
twice continuously differentiable and $f^{\prime\prime}(x)$ is
uniformly bounded on the subintervals $(i/m,(i+1)/m)$ for
$i=0,1,\dots,m-1$, it is easy to see that
$$
\int_0^1f(x)\,dx\,-\,\mbox{S}_N(f)\,=\,
O\left( \frac1{N^2}\right).
$$
We define $N$ such that $N^{-2}$ is of order $\e$. 

We now apply $\mbox{QS}_N(f)$ algorithm to compute 
an $\Theta(\e)$-approximation with
probability~$\tfrac34$ to $\mbox{S}_N(f)$, or, equivalently to
$\int_0^1f(x)\,dx$. To do it, we need to use the bit query $Q_f$ for
the function $f$, 
although so far we assumed that we can use only bit queries 
$Q_q$ for the functions~$q$ from $\Q$. 
This problem is resolved in Section 2 of
\cite{H03} where it is shown that algorithms using the bit query $Q_f$ 
can be simulated by algorithms using bit queries $Q_q$ 
at the expense of multiplying the number of bit queries by a factor of
$2$.

{}From this and (\ref{157}) with $M=O(m^{-2})$, we conclude that its is
enough to perform of order $\min(\e^{-1/2},m^{-2}\e^{-1})$ bit
queries, $\min(\e^{-1/2},m^{-2}\e^{-1})\log\,\e^{-1}$ quantum
operations, and using of order $\log\,\e^{-1/2}$ qubits. 

We finally approximate $\l(q)$ by the following algorithm
\begin{equation}\label{158}
\phi(q)\,=\,\l_{\bar q}\,+\, {\rm QS}_N(f).
\end{equation} 
This algorithm differs from the randomized algorithm
of Section 4 since we now apply the $\mbox{QS}_N$ quantum algorithm instead
of Monte Carlo to approximate $\int_0^1f(x)\,dx$. Its error is clearly
of the form
\begin{equation*}
e^{\textrm{\rm bit-quant}}(\phi,T)\,=\, 
O\left(\delta_1+m^{-2}\delta_2+m^{-4}+\e\right).
\end{equation*}
To guarantee that this error is at most $\e$, we take
$$
\delta_1=\Theta(\e),\ \ m=\Theta(\e^{-1/3}),\ \ k=\Theta(\e^{-1/6})\ \ 
\mbox{and}\ \  \delta_2=\Theta(\e^{1/3}).
$$
Using the cost analysis of Section 4 and the results of this section,
we conclude the following theorem. 
\vskip 1pc
\begin{thm}
The Sturm-Liouville eigenvalue problem can be solved in the quantum
setting with bit queries by the algorithm $\phi$ defined by
(\ref{158}). This algorithm approximates the smallest eigenvalue
$\l(q)$ with error at most $\e$ and probability $\tfrac34$ using of
order
\begin{itemize}
\item $\e^{-1/3}$ bit queries and  function values,
\item $\e^{-1/3}\,\log\,\e^{-1}$ quantum operations,
\item $\e^{-1/2}\,\log\,\e^{-1}$ classical operations,
\item $\log\,\e^{-1}$ qubits. 
\end{itemize}
Furthermore,
\begin{equation*}
n^{\textrm{\rm bit-query}}\,=\,\Theta(\e^{-1/3}),
\end{equation*}
and
\begin{equation*}
\Omega(\cc_{{\rm bit}}\,\e^{-1/3})\,=\, {\rm comp}^{\textrm{\rm bit-query}}(\e)\,=\,
O\left((\cc+\cc_{{\rm bit}})\,\e^{-1/3}\,+\, \e^{-1/2}\log\,\e^{-1}\right).
\end{equation*}
\end{thm}

Hence, we have a sharp bound of order $\e^{-1/3}$
on the number of bit queries
whereas the upper bound on the total cost depends, as in the worst
case and randomized settings, on $\e^{-1/2}\log\,\e^{-1}$, 
which is the cost of classical computations.

\subsection{Power Queries}

In this subsection we study {\em power} queries.
We formally define them as follows. 
For some problems, a quantum algorithm 
can be written in the form
\begin{equation}
|\psi\rangle\,:=\,U_m\widetilde W_mU_{m-1}\widetilde W_{m-1}\cdots U_1
\widetilde W_1U_0 |\psi_0\rangle.
\label{eq:qac}
\end{equation}
Here $U_1,\dots,U_m$ denote unitary matrices independent
of the function $q$, just as before, whereas the 
unitary matrices $\widetilde W_j$ are of the form controlled-$W_j$,
see \cite[p.~178]{nielsen}. That is,
$W_j=W^{p_j}$ for an $n\times n$ unitary matrix $W$ that depends on the input 
of the computational problem, and for some non-negative integers
$p_j$, $j=1,2,\dots,m$. Without loss of generality we assume that $n$
is a power of two. Let $\{|y_k\rangle\}$ be orthonormalized
eigenvectors of $W$, $W|y_k\rangle=\alpha_k|y_k\rangle$ with the
corresponding eigenvalue $\alpha_k$, where $|\alpha_k|=1$
and $\alpha_k=e^{\mathrm{i}\l_k}$ with $\l_k\in[0,2\pi)$ 
for $k=1,2,\dots,n$. 
For the unit vectors $|x_{\ell}\rangle=
\a_{\ell}|0\rangle+\beta_{\ell}|1\rangle\in \complex^2$,
$\ell=1,2,\dots,r$, the quantum query $\widetilde W_j$ is defined as 
\begin{equation}\label{control}
\widetilde W_j\,
|x_1\rangle|x_2\rangle\cdots|x_r\rangle|y_k\rangle\,=\,
|x_1\rangle|\cdots|x_{j-1}\rangle\bigg(\a_j|0\rangle+\beta_j
e^{\mathrm{i}\gamma p_j\l_k}|1\rangle\bigg)|x_{j+1}\rangle\cdots
|x_r\rangle|y_k\rangle.
\end{equation}
Hence, $\widetilde W_j$ is a $2^{\nu}\times 2^{\nu}$ unitary matrix with
$\nu=r+\log\,n$. We stress that the exponent $p_j$ only affects
the power of the complex number $e^{\mathrm{i}\gamma \l_k}$. 

We call $\widetilde W_j$  a {\it power} query since
they are derived from powers of $W$. 
Power queries have been successfully used for a number of
problems, see again \cite{nielsen}, including the phase estimation 
problem that will be discussed in the next section.
The phase estimation algorithm, see \cite{cleve,nielsen},
is at the core of many quantum algorithms. It
plays a central role in the fast quantum algorithms for factoring 
and discrete logarithms of Shor \cite{shor}. We stress that for Shor's
algorithm, power queries can be implemented by a number of
elementary quantum gates that is polylog  in $n$. The phase estimation
algorithm 
approximates an eigenvalue of a unitary operator $W$ using the
corresponding eigenvector, or its approximation, as part of the initial state. 
The powers of $W$ are defined by
$p_i=2^{i-1}$. Therefore, phase estimation uses 
queries with $W_1=W$, $W_2=W^{2}$, $W_3=W^{2^2}$, $\dots$, $W_m=W^{2^{m-1}}$.
It is typically assumed, see \cite{cleve}, that
we do not explicitly know $W$ but we are given quantum devices
that perform controlled-$W$, controlled-$W^2$, controlled-$W^{2^2}$, and
so on.

For the Sturm-Liouville eigenvalue problem, as well as for problems
studied in \cite{PW04}, we will use the matrix
\begin{equation}\label{eq:W}
W\,=\,\exp\left(\mathrm{i}\c M_q\right)\quad\mbox{with}\ 
\mathrm{i}=\sqrt{-1}\ \mbox{and a positive}\ \c,
\end{equation}
where the $n\times n$ matrix $M_q$ was introduced in Section 3.2 
as a discretization of the differential operator $\L_q$. 
The matrix $W$ is unitary since $M_q$ is symmetric.

For the $\widetilde W_j$ with the matrix $W$ of (\ref{eq:W}) we modify 
the query definition in equation (\ref{eq:qa}) and assume, as in 
\cite[Ch.~5]{nielsen}, that for each $j$ the $\widetilde W_j$ 
is one quantum query. 
Accordingly, for algorithms that can be expressed in the form (\ref{eq:qac}),
the number of power queries is $m$, independently of the powers $p_j$.
By analogy with (\ref{eq:wperr}),
we denote their error by $e^\w(\hat\l,m)$.

Allowing quantum algorithms of the form (\ref{eq:qac}) with power
queries, we define 
the power query complexity $n^{\rm power-query}(\e)$ to be the minimal 
number of power queries required to approximate the Sturm-Liouville 
eigenvalue problem with error $\e$, i.e.,
$$
n^{\rm power-query}(\e) = \min \{ m:\; \exists\; \hat\lambda 
\ \ \mbox{such\ that}\ \  e^\w(\hat\lambda, m)\le \e \}. $$
The cost of one power query is denoted by $\cc_{{\rm power}}$.
The total complexity, $\comp^{{\rm power-query}}(\e)$, is the defined
as the minimal cost of a hybrid algorithm in the same way as for bit
queries.

We will use the phase estimation algorithm as a basic
module for approximating the smallest eigenvalue $\l(q)$. As shown by
Abrams and Lloyd~\cite{abrams}, the phase
estimation algorithms can also be used if a good approximation of the 
eigenvector corresponding to the smallest eigenvalue is known. Such an
approximation is obtained by the algorithm of Jaksch and 
Papageorgiou~\cite{jaksch}. 
Combining these algorithms, we obtain the quantum 
algorithm that computes the smallest eigenvalue with error
$\e$ and probability $\tfrac34$ 
using $\Theta(\log\, \e^{-1})$ power queries, and  
$\Theta(\log\,\e^{-1})$ qubits.

For the sake of completeness, we review
the phase estimation problem and algorithm, the results of Abrams and 
Lloyd and the results of Jaksch and Papageorgiou 
in the next subsections. 

\subsection{Phase Estimation}

Consider $W$ defined by
(\ref{eq:W}) with $\gamma=\tfrac12$, i.e.,
$$
W\,=\,\exp\left(\frac12\,\mathrm{i}\,M_q\right). 
$$
The eigenvalues of $W$ are $e^{\mathrm{i}\l_j(M_q)/2}$ with
$\l_j(M_q)$ being the eigenvalues of the $n\times n$ matrix $M_q$
and $n$ is assumed to be a power of two. 
These eigenvalues can be written as $e^{2\pi \mathrm{i}
\varphi_j}$, where
$$
\varphi_j\,=\,\varphi_j(M_q)\,=\,\frac1{4\pi}\,\l_j(M_q)
$$
are called {\it phases}. We are interested in estimating the smallest
phase $\varphi_1(M_q)$, which belongs to $(0,1)$ since
$\l_1(M_q)\in[\pi^2,\pi^2+1]$. For convenience, we renumber and
normalize the eigenvectors of $M_q$, and also of $W$, as 
$$
|y_j\rangle\,=\,\sqrt{n}\,|z_{j+1}(M_q)\rangle,
$$ 
for $j=0,1,\dots,n-1$. We will use $\{|y_j\rangle\}$ as the
orthonormal basis of the space. 

Phase estimation, see \cite[Section 5.2]{nielsen},
is a quantum algorithm that approximates  
the phase $\varphi_1(M_q)$. Note that to compute an $\e$-approximation
of $\l_1(M_q)$, it is enough to compute an $\e/(4\pi)$-approximation
of $\varphi_1(M_q)$. The original 
phase estimation algorithm has been derived for the initial state
$|0^{\otimes m}\rangle|y_0\rangle$, where $m$ is related to
the accuracy and will be determined later, and 
$|y_0\rangle=|y_0(M_q)\rangle$ 
is the eigenvector of the matrix $M_q$ corresponding to the smallest 
eigenvalue $\l_1(M_q)$. Abrams and Lloyd \cite{abrams} showed 
that phase estimation can still be
used if the eigenvector $|y_0\rangle$ is replaced by a {\it good}
approximation $|\psi_0\rangle$ as the initial state. 

More precisely, expanding $\ket{\psi_0}$ in the basis of the eigenvectors 
$|y_j\rangle$, the initial state takes the form  
\begin{equation*}
\ket{0}^{\otimes m}\ket{\psi_0} = \ket{0}^{\otimes m}
\sum_{j=0}^{n-1} d_j\ket{y_j}.
\end{equation*}
Using $m$ Hadamard gates, we place the first register in an equal 
superposition, which gives the state 
\begin{equation*}
|\psi_1\rangle\,=\,
\frac{1}{\sqrt{2^m}}\sum_{x_1=0}^1\sum_{x_2=0}^1\cdots
\sum_{x_m=0}^1\ket{x_1}\ket{x_2}\cdots\ket{x_m}\sum_{j=0}^{n-1}d_j\ket{y_j}.
\end{equation*}
We now apply the controlled quantum gates, see (\ref{control}), to
create the state  
\begin{eqnarray*}
|\psi_2\rangle\,&=&\,\widetilde W_{2^{m-1}}\widetilde
 W_{2^{m-2}}\cdots \widetilde W_{2^0}\,|\psi_1\rangle\\
&=&\,\frac{1}{\sqrt{2^m}}\sum_{j=0}^{n-1}d_j\ket{\eta_j}|y_j\rangle
\end{eqnarray*}
with
\begin{eqnarray*}
|\eta_{j}\rangle\,&=&\,
\bigg(|0\rangle+e^{2\pi\mathrm{i}\varphi_j}|1\rangle\bigg)\otimes
\bigg(|0\rangle+e^{2\pi\mathrm{i}2\varphi_j}|1\rangle\bigg)\otimes\cdots
\otimes
\bigg(|0\rangle+e^{2\pi\mathrm{i}2^{m-1}\varphi_j}|1\rangle\bigg)\\
&=&\,\sum_{x_1=0}^1\sum_{x_2=0}^1\cdots\sum_{x_m=0}^1
e^{2\pi\mathrm{i}(x_12^0+x_22^1+\cdots x_m2^{m-1})\varphi_j}|x_1\rangle
|x_2\rangle\cdots|x_m\rangle\\
&=&\,\sum_{\ell=0}^{2^m-1}
e^{2\pi\,\mathrm{i}\,\ell\,\varphi_j}|\ell\rangle,
\end{eqnarray*}
see also \cite[p.~222]{nielsen}. Hence,
$$
|\psi_2\rangle\,=\,\frac{1}{\sqrt{2^m}}\sum_{j=0}^{n-1}d_j
\left(\sum_{\ell=0}^{2^m-1}e^{2\pi\,\mathrm{i}\,\ell\,\varphi_j}
|\ell\rangle\right)
|y_j\rangle.
$$
The inverse Fourier transform performed on the
first register creates the state
\begin{equation*}
\sum_{j=0}^{n-1} d_j \left( \sum_{\ell=0}^{2^m-1} g(\varphi_j,\ell) 
\ket{\ell} \right) \ket{y_j}, 
\end{equation*}
where 
\begin{equation*}
g(\varphi_j,\ell) = \left\{ \begin{array}{ll} \frac{\sin
(\pi(2^m\varphi_j -\ell))
e^{\pi\mathrm{i}(\varphi_j - \ell 2^{-m})(2^m-1)}} {2^m\sin(\pi 
(\varphi_j-\ell2^{-m}))}
& \quad \mbox{if}\ \  \varphi_j \neq 2^{-m}\ell, \\
1, & \quad \mbox{if} \ \ \varphi_j = 2^{-m}\ell.
\end{array} \right.
\end{equation*}
A measurement of the first register produces the outcome $j$ with probability
\begin{equation*} 
p_j = \sum_{\ell=0}^{n-1} |d_{\ell}|^2|g(\varphi_{\ell},j)|^2,
\end{equation*}
and the second register collapses to the state
\begin{equation*}
\sum_{\ell=0}^{n-1} \frac{d_{\ell} g(\varphi_{\ell},j)}{\sqrt{p_j}} 
\ket{y_{\ell}}. 
\end{equation*}
The quantity 
$$
\Delta(\phi_0,\phi_1)\,=\,\min_{x \in \integers}\{|x + \phi_1 -
\phi_0|\}
\qquad \mbox{for}\ \ \phi_0, \phi_1 \in \reals
$$
is defined in \cite{brassard} and is 
the fractional part of the distance between two phases $\phi_0$ and
$\phi_1$. It is used to derive the relationship between the 
approximation error and the success probability. 
A measurement of the first register produces an outcome 
from the set 
$$
\mathcal{G}_k\,=\, \{ j: \Delta(j/2^b,\varphi_1(M_q)) \leq k/2^m\, \},
$$
where $k>1$, with probability
\begin{equation*}
\Pr (\mathcal{G}_k)\,=\, 
        \sum_{j \in \mathcal{G}_k} \sum_{\ell=0}^{n-1} |d_{\ell} 
g(\varphi_{\ell},j)|^2 
\,\ge\, |d|^2 \sum_{j \in \mathcal{G}_k}  |g(\varphi_1(M_q) ,j)|^2
\,\ge\, |d|^2 - \frac{|d|^2}{2(k-1)},
\end{equation*}
where $d=\langle y_0 | \psi_0\rangle $.
For $k=1$ the probability that 
\begin{equation}\label{k=1}
\Delta(j/2^m,\varphi_1(M_q))\,\leq\, 2^{-m}
\quad\mbox{is bounded from below by}\ \frac{8}{\pi^2}|d|^2.
\end{equation} 
The proof of the probability bounds can be found in 
\cite{brassard,nielsen}. 
Using this fact, the authors of~\cite{abrams} conclude that
as long as $|d|^2$ is {\it large enough} or, equivalently, $\ket{\psi_0}$
is {\it close enough} to $\ket{y_0}$ then phase estimation can be used
to approximate the phase $\varphi_1(M_q)$ with probability close to
$8/\pi^2=0.81\dots$.

We stress that the phase estimation algorithm uses $m$ power
queries. In addition to the cost of the queries there is a quantum 
operations cost proportional to at most $m^2$, which is an upper bound on 
the cost of the quantum 
inverse Fourier transform, see \cite[Section 5.2]{nielsen}. 

\subsection{Eigenvalue and Eigenvector Approximation}

The results of Jaksch and Papageorgiou \cite{jaksch} 
can be applied to efficiently construct a good approximate eigenvector 
when $W=e^{\frac{\mathrm{i}}{2}M_q}$ as in the previous subsection. 

The matrix $M_q=M_q^{(n)}$ has been derived from the discretization 
of the operator $\L_q$ with mesh size $h_n= (n+1)^{-1}$. Its
eigenvectors are also eigenvectors of $W=W^{(n)}$, and we denote them
here by $\ket{y_j^{(n)}}$, where $j=0,1,\dots,n-1$. 
We want to approximate $\l_1(M_q^{(n)})=4\pi \varphi_1(M_q^{(n)})$ 
but we do not know the corresponding eigenvector 
$$
\ket{y^{(n)}}\,:=\,\ket{y_0^{(n)}}.
$$
The expansion of $\ket{y^{(n)}}$ in the computational basis is 
denoted by
\begin{equation}
\ket{y^{(n)}} = \sum_{j=0}^{n-1} y_j^{(n)}\ket{j},
\label{compexpansion}
\end{equation}
Recall that $u_q$ is the normalized, $\|u_q\|_{L_2}=
\left(\int_0^1 u^2(x)\, dx\right)^{1/2}=1$, eigenfunction of the differential
operator $\L_q$ that corresponds to $\l(q)$, and 
$u_q$ as well as $u_q^{\prime}$ and $u_q^{\prime\prime}$ are 
uniformly bounded, i.e., 
$\|u_q \|_\infty $, $\| u^{\prime}_q \|_\infty $ and
$\|u^{\prime\prime}_q\|_{\infty}$ are $O(1)$.

Let
$\ket{U^{(n)}}=\sum_{j=0}^{n-1} u_q((j+1)h_n) \ket{j}$ be 
the vector obtained by sampling $u_q$ at the discretization points. 
Then it is known, see \cite{gary,keller} as well as Remark 4.1, that
\begin{eqnarray}
\left\| \ket{y^{(n)}} - 
\frac{\ket{ U^{(n)} }}{\| \ket{ U^{(n)} }\|_2} \right\|_2
&=&O(h_n^2) \quad {\rm and} \label{pointwise} \\
|\l(q)-\l_1(M_q^{(n)})|&=&O(h_n^2). \nonumber
\end{eqnarray}
Consider a coarse discretization of $\L_q$ with mesh size
$h_{n_0}=(n_0+1)^{-1}$ with $n_0$ being a power of two.  
Assume that 
$$
\ket{\tilde{z}^{(n_0)}}\,=\,\sum_{j=0}^{n_0-1}\tilde{z}^{(n_0)}_j\ket{j},
\quad \|\ket{\tilde{z}^{(n_0)}}\|_2=1,
$$ 
approximates the eigenvector $\ket{y^{(n_0)}}$ that corresponds to the
smallest eigenvalue of the matrix $M_q^{(n_0)}$ such that,
\begin{equation}\label{newapprox}
\|\,\ket{\tilde{z}^{(n_0)}}-\ket{y^{(n_0)}}\,\|_2\,=\,O(n_0^{-2}).
\end{equation}
We place the vector $\ket{\tilde{z}^{(n_0)}}$ 
in a $\log\,n_0$ qubit register. As explained in Section 4.3, we can compute 
$\ket{\tilde{z}^{(n_0)}}$ on a classical computer
with cost of order $n_0^2$. 

For $n=2^s n_0$, we construct an approximation $\ket{\tilde{z}^{(n)}}$ 
of $\ket{y^{(n)}}$ by first appending $s$ qubits, all in the 
state $\ket{0}$, to $\ket{\tilde{z}^{(n_0)}}$ and then performing the  
Hadamard transformation on each one of these $s$ qubits, i.e.,
\begin{equation}
 \ket{ \tilde{z}^{(n)}} =  \ket{\tilde{z}^{(n_0)}} 
\left( \frac{\ket{0}+\ket{1}}{\sqrt{2}} \right)^{\otimes s}
=  \frac{1}{\sqrt{2^s}} 
\sum_{j=0}^{n-1} \tilde{z}_{g(j)}^{(n_0)}\,\ket{j},
\label{eq:JPalg}
\end{equation} 
where $\tilde{z}_{g(j)}^{(n_0)}$'s denote the coordinates
of $\ket{\tilde{z}^{(n_0)}}$ in the computational basis, and 
$g(j) = \lfloor j/2^s \rfloor$. 
The effect of $g$ is to replicate $2^s$ times the coordinates of 
$\ket{\tilde{z}^{(n_0)}}$. 
As in Jaksch and Papageorgiou \cite{jaksch}, we use
the vector $\ket{ \tilde{z}^{(n)} }$ as part of the input to the phase
estimation algorithm.

Let $d^{(n)}= \langle y^{(n)} | \tilde{z}^{(n)}\rangle$.
We show that $|d^{(n)}|^2$ can be made arbitrarily close to one by 
choosing a sufficiently large $n_0$. 
Hence, we can make the success probability
of the phase estimation algorithm at least equal to $\tfrac34$. 

Consider two different expansions of $\ket{\tilde{z}^{(n)}}$, 
\begin{eqnarray}
\ket{\tilde{z}^{(n)}} &=& \sum_{j=0}^{n-1} \tilde{u}_{j}^{(n)} \ket{j} 
\label{approxexpansion}\\
\ket{\tilde{z}^{(n)}} &=& \sum_{j=0}^{n-1} d_{j}^{(n)} \ket{y_j^{(n)}}. 
\label{eigenexpansion}
\end{eqnarray}
The first expansion is in the computational basis $\{|j\rangle\}$
and, by (\ref{eq:JPalg}),
$$
\tilde{u}_j^{(n)}=2^{-s/2}z_{g(j)}^{(n_0)}\qquad \mbox{for}\ j=0,1,\dots,n-1,
$$
while the second expansion is with 
respect to the eigenvectors of $M_q^{(n)}$. 
Note that $d^{(n)}=d_0^{(n)}$ and clearly $\sum_{j=0}^{n-1}|d_j^{(n)}|^2=1$. 
Equation 
(\ref{eigenexpansion}) implies 
\begin{equation} 
\ket{\tilde{z}^{(n)}} - \ket{y^{(n)}} = 
(d^{(n)} - 1) \ket{y^{(n)}} 
+ \sum_{j=1}^{n-1}d_{j}^{(n)} \ket{y_j^{(n)}}.
\end{equation} 
Taking norms on both sides we obtain 
\begin{equation}\label{errorbound}
\left| \left|\, \ket{y^{(n)}} - \ket{\tilde{z}^{(n)}}\, \right|
\right|^2_2 \,=\, |d^{(n)}-1|^2 + 
\sum_{j=1}^{n-1}|d_{j}^{(n)}|^2 \,\geq\,
\sum_{j=1}^{n-1}|d_j^{(n)}|^2 \,=\, 1-|d^{(n)}|^2.
\end{equation}

We now bound the left hand side of (\ref{errorbound}) from above. 
Using the expression (\ref{compexpansion}) for
$\ket{y^{(n)}}$ and the definition of $\ket{\tilde{z}^{(n)}}$, see
(\ref{eq:JPalg}), (\ref{approxexpansion}), we have
\begin{eqnarray*}
\left\| \ket{y^{(n)}} - \ket{\tilde{z}^{(n)}} \right\|^2_2 &=&
\sum_{j=0}^{n-1} |y_j^{(n)} - 2^{-s/2}z_{g(j)}^{(n_0)}|^2 \\
&=&
\sum_{j=0}^{n-1} \Biggl| \frac{u_q((j+1)h_{n})}{\| \ket{U^{(n)}} \|_2} \Biggr. 
- \frac{u_q((g(j)+1)h_{n_0})}{\sqrt{2^s} \| \ket{U^{(n_0)}} \|_2}  + 
\Delta_{j}^{(n)} - \frac{\Delta_{g(j)}^{(n_0)}}{\sqrt{2^s}} \Biggl. \Biggr|^2,
\end{eqnarray*}
where, by (\ref{pointwise}) and (\ref{newapprox}), we have
$$
\sum_{j=0}^{n-1} |\Delta_{j}^{(n)}|^2 = O(h_n^{4})\qquad\mbox{and}\qquad 
\sum_{j=0}^{n-1} |\Delta_{g(j)}^{(n_0)}|^2 = 2^s O(h_{n_0}^{4}).
$$
Applying the triangle inequality, we get
\begin{equation}
\left\|\ket{y^{(n)}} - \ket{\tilde{z}^{(n)}} \right\|_2\,\leq\, 
\left(\sum_{j=0}^{n-1} \left| \frac{u_q((j+1)h_{n})}
{\| \ket{U^{(n)}} \|_2}  \right. \right. 
- \left. \left.  \frac{u_q((g(j)+1)h_{n_0})}{\sqrt{2^s} \| 
\ket{U^{(n_0)}} \|_2}  \right|^2 \right)^{1/2} + O(h_{n_0}^2).   
\label{sumestimate}
\end{equation}
The definition of $\ket{U^{(n)}}$ and the fact that 
the derivative of $u_q$ is Lipschitz\footnote{
A function $f:[0,1]\to\reals$ is Lipschitz if there is a number $L\ge 0$
such that $|f(x)-f(y)|\le L|x-y|$ for all $x,y\in[0,1]$.}
with the uniform Lipschitz constant
imply that $\| \ket{U^{(n)}} \|_2 = \sqrt{n}(1+O(h_n))$. Hence, the square of
the term in the parentheses above is equal to 
\begin{equation}
\frac{1}{n} \sum_{j=0}^{n-1} | u_q((j+1)h_{n}) (1 + O(h_n))
-  u_q((g(j)+1)h_{n_0}) (1 + O(h_{n_0})) |^2. 
\label{uk}
\end{equation}  
Since $u_q$ is continuous with a bounded first derivative, we have that
\begin{equation}
u_q(x_{2,j}) = u_q(x_{1,j}) + O(|x_{2,j}-x_{1,j}|), 
\label{meanvalue} 
\end{equation}
where $x_{2,j}=(j+1)h_n$ and $x_{1,j}=(g(j)+1)h_{n_0}$, 
$j=0,1,\ldots,n-1$. Let $\lfloor j/2^s\rfloor=j/2^s-\a$ with
$\a\in(0,1)$.
Then 
\begin{eqnarray*}
|x_{2,j}-x_{1,j}|\,&=&\,\left|\frac{j+1}{2^sn_0+1}-\frac{j/2^s+1-\a}{n_0+1}
\right|\\
&=&\,j\,\frac{2^s-1}{(2^sn_0+1)2^s(n_0+1)}+O(h_{n_0})\,=\, O(h_{n_0}).
\end{eqnarray*}
Using (\ref{uk}), (\ref{meanvalue}) and the triangle inequality, 
we obtain from (\ref{sumestimate}) that
$$
\left\| \ket{y^{(n)}} - \ket{\tilde{z}^{(n)}} \right\|_2\,=\,
O(h_{n_0})
\,\le\,\frac{c}{n_0+1} 
$$
for some positive number $c$ independent of $n$ and $n_0$. 
Combining this with (\ref{errorbound}) we finally conclude that
\begin{equation}\label{failure}
|d^{(n)}|^2\,\ge\,1\, -\,\frac{c^2}{(n_0+1)^2}
\end{equation} 
and $d^{(n)}$ can be made arbitrarily close to one by taking a
sufficiently large $n_0$. 

\subsection{Quantum Algorithm for the Smallest Eigenvalue}

We combine the results of the previous two subsections
to derive a quantum algorithm for computing an $\e$-approximation of
the smallest eigenvalue with probability $\tfrac34$. 
 
We choose the parameters for the phase estimation algorithm. 
Without loss of generality we assume that $\e^{-1}$ is an even power of $2$,
that is $\e^{-1}=2^{m}$ with an even $m$. We set $n=\e^{-1/2}=2^{m/2}$
and we will be working with the matrix $M_q^{(n)}$. The index $n_0=2^{k_0}$
is chosen as the smallest power of two for which
\begin{equation}
\frac8{\pi^2}\,\left(1-\frac{c^2}{(n_0+1)^2}\right)\,\ge\,\frac34,
\label{eq:n0}
\end{equation}
where the number $c$ is from (\ref{failure}). Clearly, $n_0=O(1)$. 
Without loss of generality we assume that $\tfrac12m>k_0=\log\,n_0$, 
i.e., we
assume that $\e$ is sufficiently small. We finally set
$s=\tfrac12m-k_0$. We then compute $\ket{\tilde{z}^{(n_0)}}$ 
on a classical computer as in Section 4.3 with cost $O(1)$ function
values and operations.  

We run the phase estimation algorithm for the matrix
$W=e^{\frac{\mathrm{i}}2M_q^{(n)}}$ with the initial state, see
(\ref{eq:JPalg}), 
$$
\ket{0}^{\otimes m}\ket{\tilde{z}^{(n)}}\,=\ket{0}^{\otimes m}
\ket{\tilde{z}^{(n_0)}}
\left( \frac{\ket{0}+\ket{1}}{\sqrt{2}} \right)^{\otimes s}. 
$$  
Let $j$ be the outcome of the phase estimation algorithm. We finally
compute
$$
\bar \l_j\,=\,4\pi\,j\,2^{-m}
$$
as an approximation of the smallest eigenvalue $\l(q)$. We have
\begin{eqnarray*}
\bar \l_j\,-\, \l(q)\,&=&\,\bar \l_j\,-\,\l_1(M_q^{(n)})\,+\,
\l_1(M_q^{(n)})\,-\,\l(q)\\
&=&\,4\pi\,\left(\frac{j}{2^m}-\varphi_1(M_q^{(n)})\right)\,+\,O(\e).
\end{eqnarray*}

{}From (\ref{k=1}) we know that
$$
\left(\frac{j}{2^m}-\varphi_1(M_q^{(n)})\right)\,\le\,\e
\quad
\mbox{with probability}\ \frac8{\pi^2}\,|d^{(n)}|^2.
$$
By (\ref{failure}) and the definition of $n_0$ we have
$$
\frac8{\pi^2}|d^{(n)}|^2\,\ge\,
\frac8{\pi^2}\left(1-\frac{c^2}{(n_0+1)^2}\right)\,\ge\,\frac34.  
$$
Hence,
$$
|\bar \l_j\,-\,\l(q)|\,=\,O(\e)
\quad
\mbox{with probability at least}\ \frac34.
$$
The computation of $\bar \l_j$ requires
$$
m+k_0+s\,=\,\tfrac32\,m\,=\,\tfrac32\,\log\,\e^{-1}
$$
qubits, $m=\log\,\e^{-1}$ power queries, plus
a number of quantum operations proportional to $m^2=\log^2\e^{-1}$. 
This yields $n^{\textrm{power-query}}(\e)=O(\log\,\e^{-1})$.
A lower bound on $n^{\textrm{power-query}}(\e)$ of the same order
is proved in \cite{Bessen}. Hence,
\begin{equation*}
 n^{\textrm{power-query}}(\e)\,=\,\Theta(\log\,\e^{-1}).
\end{equation*}

We summarize the results of this section in the following theorem.

\begin{thm} 
The Sturm-Liouville eigenvalue problem can be solved
in the quantum setting with power queries by the phase estimation algorithm
applied to the discretized matrix of the differential operator $\L_q$ 
with the initial state given as an approximate 
eigenvector computed by the Jaksch and Papageorgiou algorithm.
This quantum algorithm approximates the smallest
eigenvalue $\l(q)$ with error $\e$ and probability $\tfrac34$
using 
\begin{itemize}
\item 
$\tfrac32\log\,\e^{-1}+O(1)\ \ \mbox{power queries}$, 
\item $O(1)$ function values and classical operations, 
\item 
$O(\log^2\e^{-1})\ \ \mbox{quantum operations besides the power queries}$, and 
\item
$\tfrac32\log\,\e^{-1}\,+O(1)\ \  \mbox{qubits}$. 
\end{itemize}
Furthermore, 
\begin{equation*}
n^{\textrm{\rm power-query}}\,=\,\Theta(\log\,\e^{-1}),
\end{equation*}
and
\begin{equation*}
\Omega(\cc_{{\rm power}}\,\log\,\e^{-1})\,=\, 
{\rm comp}^{\textrm{\rm power-query}}(\e)\,=\,
O\left(\cc_{{\rm power}}\,\log\,\e^{-1}\,+\, \cc\,+
\log^2\e^{-1}\right).
\end{equation*}
\end{thm}

\subsection{Qubit Complexity}

In this section we address the qubit complexity, $\cb(\e)$, which is
defined as the minimal number of qubits required 
to approximate the smallest eigenvalue with error $\e$ and 
probability~$\tfrac34$ by quantum algorithms of the form
(\ref{eq:qa}). Clearly, $\cb(\e)$ is upper bounded by
$\frac32\log\,\e^{-1}+O(1)$ since that many qubits are used by the
phase estimation algorithm of Section~5.5. 
Observe that the cost of the classical algorithm
computing $\ket{\tilde{z}^{(n_0)}}$ as well as its quantum simulation
\cite[p.~189-193]{nielsen}
is constant since $n_0$ is bounded by a constant due to (\ref{eq:n0}).

We turn to a lower bound on
$\cb(\e)$.  Based on the results obtained in
this paper, it is easy to see that the number of qubits necessary to
solve our problem must be proportional at least to roughly
$\tfrac12\log\,\e^{-1}$. Indeed, assume that there is
a quantum algorithm of the form (\ref{eq:qa})
that computes $\l(q)$ with error~$\e$ and
probability $\tfrac34$, and uses
$k(\e)$ qubits. This algorithm can use arbitrary quantum queries,
assuming that each quantum query is based on at most $2^{k(\e)}$ function 
evaluations of~$q$. Note that this holds for bit queries, as well as for the
power queries studied in this paper. 
Then such an algorithm can be simulated by a classical
algorithm that uses at most $2^{k(\e)}$ function evaluations of $q$. 
{}From Theorem 3.2 we know that $2^{k(\e)}=\Omega(\e^{-1/2})$
and therefore $k(\e)\ge\tfrac12\log\,\e^{-1}+\Omega(1)$.
Hence, the qubit complexity is lower bounded by 
$\tfrac12\log\,\e^{-1}+\Omega(1)$. This 
proves the following theorem.
\begin{thm}
The qubit complexity of the Sturm-Liouville eigenvalue problem
in the quantum setting with bit or power queries is bounded by
$$
\tfrac12\,\log\,\e^{-1}\,+\,O(1)\,\le\,\cb(\e)\,\le\,
\tfrac32\,\log\,\e^{-1}\,+\,O(1).
$$ 
\end{thm} 

\section*{Acknowledgments}
\vskip 1pc

This research has been supported in part by 
the National Science Foundation, the Defense Advanced Research
Projects Agency (DARPA), and the Air Force  
Research Laboratory. 

We are grateful for valuable comments from Stefan Heinrich, Marek Kwas,
Joseph F. Traub and Arthur G. Werschulz.

\end{document}